\documentclass[11pt]{article}
\usepackage[utf8]{inputenc}
\usepackage[margin=1in]{geometry}

\usepackage{microtype}
\usepackage{graphicx}
\usepackage{subfigure}
\usepackage{booktabs}
\usepackage{natbib}
\setcitestyle{open={(},close={)}}

\usepackage{setspace}

\usepackage{amsmath}
\usepackage{amssymb,amsmath,amsthm}

\newtheorem{assumption}{Assumption}
\usepackage{mathtools}
\usepackage{amsthm}
\usepackage{algorithmic}
\usepackage{graphicx}
\usepackage{textcomp}
\usepackage{xcolor}
\usepackage{amsfonts}
\usepackage{epsfig}
\usepackage{algorithm}
\usepackage{wrapfig}
\usepackage{balance}
\usepackage{url}
\usepackage{mathtools}
\usepackage{natbib}
\usepackage{multirow}
\usepackage{soul,comment}

\usepackage[textsize=tiny]{todonotes}

\usepackage[colorlinks=true,citecolor=blue]{hyperref}

\usepackage{amsmath,amsthm,amsfonts,amssymb,mathdots,array,mathrsfs,bm,bbm,stmaryrd,graphicx,subfigure,xcolor}

\usepackage{breakcites}

\usepackage[T1]{fontenc}
\usepackage{enumerate}
\usepackage{inputenc}

\usepackage{graphicx} 
\usepackage{subfigure}

\usepackage{booktabs,balance}
\usepackage{rotating}
\usepackage{boldline}
\usepackage{makecell}
\usepackage{multirow}
\usepackage{balance}

\usepackage{tikz}

\newcommand{\bsmat}{\begin{bmatrix} }
\newcommand{\esmat}{\end{bmatrix} }

\usepackage{environ}
\NewEnviron{smallequation}{%
    \begin{equation}
    \scalebox{0.97}{$\BODY$}
    \end{equation}
    }
    \NewEnviron{smallalign}{%
    \begin{equation}
    \scalebox{0.97}{$\BODY$}
    \end{equation}
    }

\begin{document}

\title{\bf\Huge Constrained Approximate Similarity Search on Proximity Graph}

\author{\textbf{Weijie Zhao, Shulong Tan, Ping Li} \\\\
Cognitive Computing Lab\\
Baidu Research\\
10900 NE 8th St. Bellevue, WA 98004, USA\\
  \texttt{\{zhaoweijie12, tanshulong2011,  pingli98\}@gmail.com}
}

\date{\vspace{0in}}
\maketitle

\begin{abstract}\vspace{0in}
\noindent Search engines and recommendation systems are built to efficiently display relevant information from those massive amounts of candidates. Typically a three-stage mechanism is  employed in those systems: (i) a small collection of items are first retrieved by (e.g.,) approximate near neighbor search algorithms; (ii) then a collection of constraints are applied on the retrieved items; (iii) a fine-grained ranking neural network is employed to determine the final recommendation. We observe a major defect of the original three-stage pipeline: Although we only target to retrieve $k$ vectors in the final recommendation, we have to preset a sufficiently large $s$ ($s > k$) for each query, and ``hope'' the number of survived vectors after the filtering is not smaller than $k$. That is, at least $k$ vectors in the $s$ similar candidates satisfy the query constraints.

\vspace{0.1in}

\noindent In this paper, we investigate this \textbf{constrained similarity search} problem and attempt to merge the similarity search stage and the filtering stage into one single search operation. We introduce \textbf{AIRSHIP}, a system that integrates a user-defined function filtering into the similarity search framework. The proposed system does not need to build extra indices nor require prior knowledge of the query constraints.
We propose three optimization strategies: (1) starting point selection, (2) multi-direction search, and (3) biased priority queue selection.
Our first starting point selection optimization is to locate good starting points for the graph searching algorithm. We would like to start the graph search inside a cluster of satisfied vectors. In this scenario, a small number of satisfied vectors can be retrieved along the path to the query vector.
Then, we propose a multi-direction search optimization that enables multi-direction searching. Compared with the single priority queue used in the original graph search, we have two priority queues to keep search candidates: on stores the satisfied vectors and the other maintains remaining unsatisfied vectors. When both priority queues are not empty, we choose each queue with a weight ratio, $alter\_ratio$.
On top of that, we propose a biased priority queue selection to adaptively select the queue for searching: when the top candidate from the satisfied vector queue is better than the one from the other queue, we override the $alter\_ratio$ restriction and select the satisfied vector queue.
All three optimizations improve the similarity search performance. The proposed algorithm has a hyper-parameter \textit{alter\_ratio} that is data/query dependent. We present an estimation method to adaptively choose the hyper-parameter.
Experimental evaluations on both synthetic and real data confirm the effectiveness of the proposed \textbf{AIRSHIP} algorithm.

\vspace{0.1in}

\noindent We focus on constrained graph-based approximate near neighbor (ANN) search in this study, in part because graph-based ANN is known to achieve excellent performance. We believe it is also possible to develop constrained hashing-based ANN or constrained quantization-based ANN.

\end{abstract}

\newpage

\section{Introduction}\label{sec:intro}

The task of searching for similar items is the standard routine in numerous industrial applications as well as research problems in computer science and other fields. Finding the exact near (or nearest) neighbors is often expensive and typically an approximate solution would be sufficient.  The research on developing efficient algorithms for approximate near neighbor (ANN) search dated back at least to the 1970's~\citep{friedman1975algorithm,friedman1977algorithm}. In recent years,  ANN algorithms have become the crucial component in modern recommender systems~\citep{davidson2010youtube,dror2011i,fan2019mobius,kanakia2019scalable}, Among many  application domains. In this paper, we study ANN algorithms  for an important application scenario, that is, the retrieved items must satisfy certain pre-specified constraints. More specifically, we focus on graph-based ANN algorithms under constraints. We should emphasize that, although this research problem was motivated from the production needs at Baidu, we only report experiments using public datasets.

\subsection{Algorithms for Approximate Near Neighbor (ANN) Search}

Broadly speaking, most approximate similarity search methods fall in the following 5 categories:
\begin{itemize}
\item Hash-based methods. Traditionally, hashing-based methods~\citep{broder1997syntactic,indyk1998approximate,charikar2002similarity,datar2004locality,li2005using,lv2007multi,pauleve2010locality,shrivastava2012fast,shrivastava2014defense,xinan2016an,li2019sign,li2021consistent} partition the base vectors into a constant number of buckets via a specific hashing function. The chosen hashing function guarantees a collision probability is positively correlated to the specific similarity of two vectors, i.e., similar vectors have a high probability to be assigned to the same hash bucket. Relatively recently, learning to hash methods~\citep{kulis2009learning,grauman2013learning,lai2015simultaneous,wang2016learning,Dong2020Learning} become popular. Typically they learn the hashing on a sample of data so that the learned hash function can better fit the data distribution compared with the classic data-independent hashing algorithms.

\item Tree-based methods~\citep{friedman1975algorithm,friedman1977algorithm,jagadish2005idistance,cayton2008fast,ram2012maximum,curtin2013fast,ram2019revisiting} partition the high-dimensional data space into smaller regions via a decision tree/space partitioning tree. The partitioned spaces are organized within a tree: A leaf node in the tree corresponds to the finest granularity of the partitioned space, and edges between tree nodes represent the hierarchical relationship of partitioned spaces. \citet{hamilton2020mosaic} recently propose a tree-based method that builds an inverted index for each label/attributes. It is infeasible to build an index for each label combination without prior knowledge of the query constraints. Our proposed solution performs the original index and does not need to build any additional indices. Moreover, we do not need to know the query constraints beforehand.

\item Quantization-based algorithms~\citep{jegou2011product,ge2013optimized,wang2016supervised,wu2017multiscale,xu2018online,echihabi2021new} learn a code-book from base vectors and quantize the original high-dimensional vectors into a low-dimensional quantized values. The quantized value is a low-dimensional representation of the original vectors. The distances between quantized vectors approximate the distances between original vectors.

\newpage

 \item Graph-based algorithms achieve a lot of attention recently due to its superior performance to other similarity search techniques on many real-world datasets and industrial applications~\citep{hajebi2011fast,wu2014fast,malkov2014approximate,malkov2020efficient,fu2019fast,tan2019efficient,zhou2019mobius,tan2021norm,xu2022proximity}. Graph-based algorithms build a proximity graph as an index and perform a graph searching to answer the query. We investigate the constrained similarity search on the graph-based algorithm. Recently, \cite{zhao2020song} develop the first GPU-based graph-based ANN algorithm.

\item Neural ranking. To resolve issues related to proximity search in various neural models, in recent years, the subject of ``neural ranking'' has attracted increasingly more attentions~\citep{zhu2018learning,zhu2019joint,tan2020fast,zhuo2020learning,gao2020deep,tan2021fast,yu2022egm,zhao2022guitar}. Effective algorithms for neural ranking can be very desirable in practice. For example, while the standard ``two-tower'' model is popular and fairly effective, practitioners might hope to replace the simple ``cosine'' loss function with a most sophisticated deep neural nets for better retrieval accuracy. However, once we have replaced the cosine by a neural net, we can no longer use the standard ANN techniques to achieve fast ranking.

\end{itemize}

Among the above methods, hashing-based methods are typically the simplest but the performance highly depends on the data and the chosen hashing functions. Graph-based ANN methods typically outperforms other methods, often considerably. In this study, we focus on graph-based ANN methods.

\subsection{Approximate Near Neighbor Search Under Constraints}

\begin{figure}[htp]

\begin{center}
\includegraphics[width=4.8in]{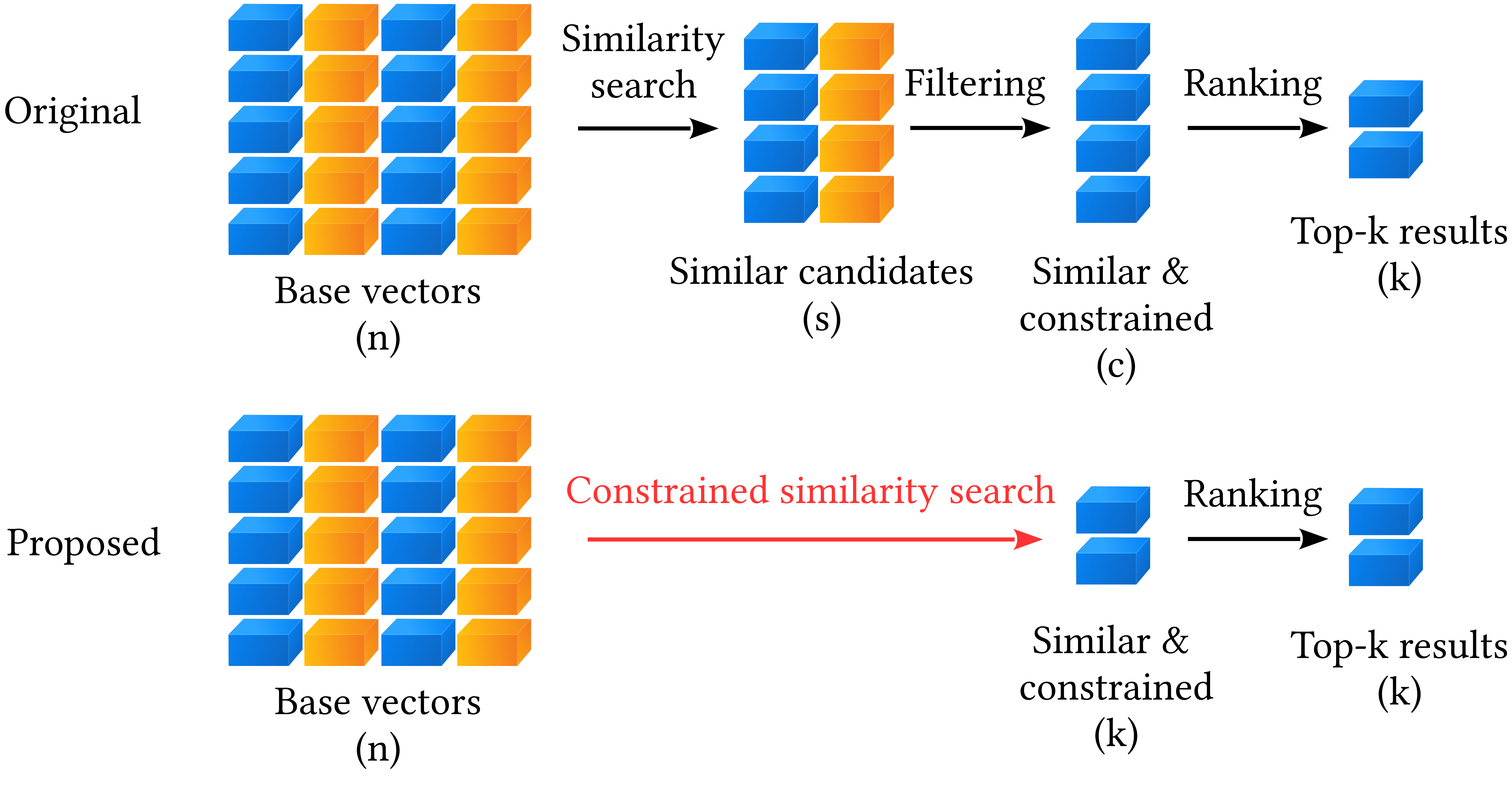}
\end{center}

\vspace{-0.4in}

\caption{The three-stage pipeline v.s our proposed constrained similarity search workflow.}
\label{fig:pipeline}
\end{figure}

In this paper, we consider a practical scenario commonly encountered in industrial applications, as illustrated in Figure~\ref{fig:pipeline}. That is, the returned search results have to satisfy certain constraints. While engineers and researchers from industry would probably agree this is a ubiquitous task in general, how to effectively solve the problem depends on the pattern of the constraints.

In Figure~\ref{fig:pipeline} (upper), we describe the usual strategy. That is, we consider there are initially $n$ vectors. For a particular query, the ANN search algorithm returns $s$ vectors (from the $n$ vectors). After applying the constraints on the $s$ vectors, the number is reduced to $c$. Finally, we apply the re-ranking algorithm (which can be either simply based on the similarity or based on the prediction results from a neural net or other learning model)  to obtain the final top-$k$ results. Obviously, with this strategy, practitioners would like to make sure that $c>k$, but this is not always easy to achieve unless we increase $s$ to be much larger than $k$. For example, when the target is to retrieve the top-1000 vectors, practitioners may have to first retrieve the top-$5000$ or even top-$10000$ vectors and then apply the constraints on the retrieved vectors. In this study,  we hope to develop better schemes to improve the efficiency.

Different from the  pipeline in the upper part of Figure~\ref{fig:pipeline}, there is another obvious strategy by simply applying the constraints on the original $n$ vectors before conducting approximate near neighbor search. If engineers know (or guess) beforehand that directly filtering on the $n$ vectors would result in only a small number of vectors, then this would be a good strategy, provided that the cost for applying the constraints on all $n$ vectors is not too high (for example, when indices have been built for those constraints).  In this paper, we do not consider this scenario.

As illustrated in the bottom part of Figure~\ref{fig:pipeline}, in this study, our goal is to directly retrieve the top-$k$ vectors which satisfy the constraints. We focus on the graph-based ANN algorithm in this paper and we believe that one can also develop, with efforts, retrieval algorithms of similar nature based on other ANN methods such as hashing-based or quantization-based ANN algorithms. We leave them for future research.

\subsection{Constrained Similarity Search}

More formally, we merge the similarity search stage and the filtering stage into one single search operation. The merging has two principal advantages:
\begin{enumerate}
\item The constrained similarity search targets to return top-$k$ similar candidates that satisfy the constraints. We do not need to estimate $s$ to ensure the search performance.
\item The similarity search algorithm takes the constraints into the consideration---additional indices and optimizations can be potentially applied to the search.
\end{enumerate}

\vspace{0.1in}
\noindent\textbf{Problem Statement.}
We have a collection of $n$ $d$-dimensional base vectors $V = \{v_{1},v_{2},\cdots,v_{n}\}$. Each vector has $m$ attributes: the attributes of the $i^{\textit{th}}$ vector are $\{a_{i1},a_{i2},\cdots,a_{im}\}$.
In addition, we are given $Q$ queries. The $j^\textit{th}$ query consists of a query vector $q_j$ and a user-defined function $f_j$ as the constraint. For each query vector $q_j$, we are required to return $k$ similar candidates ($r_{j1},r_{j2},\cdots,r_{jk}$) from base vectors and each candidate $r_{j\cdot}$ satisfies the constraint, i.e., $f_j(r_{j\cdot}) \rightarrow \texttt{true}$.

\vspace{0.1in}
\noindent
\textbf{Challenges \& approaches.}
We identify two main challenges for the constrained similarity search.
The first challenge is that we are required to efficiently identify the ``interesting region'' that contains vectors that meet the given constraints. There is no sub-linear algorithm to tackle this challenge without any assumptions to the query constraints and vector distributions: Before knowing queries, it is impractical to build indices for all possible constraints in pre-processing. On the other hand, after queries arrive, building indices for all different constraints is more time-consuming than a linear scan.
Therefore, we have one assumption in this paper: for each constraint, there are at least $p\%$ of base vectors are satisfied vectors.

Secondly, the distributions of the vectors that satisfy the constraints are unknown. We need to retrieve similar vectors from them. However, similarity search techniques require us to construct an index beforehand. In this paper, we propose an adaptive searching algorithm on proximity graph that explores the clusterness of the vector distribution---vectors which meet the constraints are very likely to cluster in real-world datasets because the vector representations are trained based on those attributes. We leverage a heuristic algorithm to determine hyper-parameters in the proposed searching algorithm.

\vspace{0.1in}
\noindent
\textbf{Why not inverted index?} We have explained this earlier but we would like to re-iterate the problem since it is a common question. Basically, for applications in which it is efficient to build indices for the constraints and the fraction of data vectors after filtering is small, then we might want to go with this straightforward approach. In real applications, however, it is often inefficient to build indices for all the (potentially combinatorial number of) constraints.  Creating this many indices is not only time-consuming but also space-intensive, because we would need to duplicate a vector each time it is captured by a query constraint. Note that the satisfied vectors for different query constraints may overlap. Hence, in this paper we seek more efficient solutions for scenarios in which the inverted index approach is not applicable.

\vspace{0.1in}
\noindent\textbf{Contributions.}
Our major contributions are summarized as follows:
\begin{itemize}
\item We introduce AIRSHIP (an acronym for AttrIbute-constRained Similarity searcH on proxImity graPh) that integrates a user-defined function filtering into the similarity search framework. AIRSHIP requires no extra indices nor prior knowledge of the query constraints.
\item We propose three optimizations for the attribute-constrained searching problem: starting point selection, multi-direction search, and biased priority queue selection.
\item We present an estimation method to choose the hyper-parameter of the searching algorithm.
\item We experimentally evaluate the proposed algorithm on both synthetic and real datasets. The empirical results confirm the effectiveness of our optimizations.
\end{itemize}

\section{AIRSHIP: Attribute-Constrained Similarity Search}\label{sec:airship}
In this section, we introduce our solutions to the attribute-restricted similarity search problem. We begin with a baseline adaption of graph-based similarity search methods. Then, we discuss the drawbacks of the baseline solution and propose our optimizations step by step. Each optimization is built on top of the previous one.

\begin{spacing}{0.8}
\begin{algorithm}[htbp]
\algsetup{linenodelimiter=.}
\caption{Vanilla Graph-Based Similarity Search}\label{alg:vanilla}
\begin{flushleft}
\textbf{Input:} A Proximity graph $G(V,E)$; a query vector $q$; a user-defined query constraint $f$; a number of output candidates $K$.\\
\textbf{Output:} Top-$K$ vectors for the query $q$ with the constraint $f$.
\end{flushleft}
\begin{algorithmic}[1]
\STATE Initialize a min-heap priority queue $\textit{pq}$; a max-heap priority queue $\textit{topk}$.\label{alg:vanilla-init-begin}
\STATE find a random starting point $S$
\STATE $\textit{d} \leftarrow \textit{dist}(S,q)$
; $\textit{pq} \leftarrow \textit{pq} \cup \{(d,S)\}$
; $\textit{visited} \leftarrow \{S\}$\label{alg:vanilla-init-end}

\WHILE{$\textit{pq} \not= \emptyset$}
    \STATE $(\textit{now\_dist},\textit{now\_idx}) \leftarrow \textit{pq}.\textit{pop\_min}()$\label{line:vanilla-extract}
    \IF{$|\textit{topk}| = K$  \textbf{and} $\textit{now\_dist} > \textit{topk}.\textit{peek\_max}()$}\label{alg:vanilla-terminate-begin}
        \STATE \textbf{break}
    \ENDIF\label{alg:vanilla-terminate-end}
    \IF{$f(now\_idx) = \textbf{true}$}\label{line:vanilla-update-begin}
        \STATE $\textit{topk} \leftarrow \textit{topk} \cup \{\textit{now\_dist},\textit{now\_idx}\}$
        \IF{$|\textit{topk}| > K$}
            \STATE $\textit{topk}$.\textit{pop\_max}()
        \ENDIF
    \ENDIF\label{line:vanilla-update-end}
    \FOR{\textbf{each} \textit{next\_idx} $\in$ neighbors of \textit{now\_idx} in $G$}\label{line:vanilla-search-begin}
        \IF{$\textit{next\_idx} \not\in \textit{visited}$}
            \STATE $\textit{d} \leftarrow \textit{dist}(\textit{next\_idx},q)$
            \STATE $\textit{pq} \leftarrow \textit{pq} \cup \{ (d,\textit{next\_idx})\}$
            \STATE $\textit{visited} \leftarrow \textit{visited} \cup \{\textit{next\_idx}\}$
        \ENDIF
    \ENDFOR\label{line:vanilla-search-end}
\ENDWHILE
\RETURN \textit{topk}
\end{algorithmic}
\end{algorithm}
\end{spacing}

\subsection{Vanilla Graph-Based Similarity Search}
Algorithm~\ref{alg:vanilla} illustrates the workflow of the vanilla graph-based similarity search.
From Lines~\ref{alg:vanilla-init-begin}-\ref{alg:vanilla-init-end}, a random starting point $S$ is selected and base data structures are initialized.
In each iteration, the vector that is closest to the query point is extracted from the priority queue (Line~\ref{line:vanilla-extract}).
The algorithm is terminated when we have obtained $K$ candidates and the current searched vector is worse than the worst vector in \textit{topk} (Lines~\ref{alg:vanilla-terminate-begin}-\ref{alg:vanilla-terminate-end}).
The only difference to the original graph-based similarity search algorithm locates in Lines~\ref{line:vanilla-update-begin}-\ref{line:vanilla-update-end}: We only update the top-$K$ candidates when the current searched vector satisfies the query constraint. After that, Lines~\ref{line:vanilla-search-begin}-\ref{line:vanilla-search-end} iterate over all unvisited neighbors of \textit{now\_idx}, add them to the priority queue, and mark them as visited.

\begin{figure*}[htp]
\includegraphics[width=1.0\textwidth]{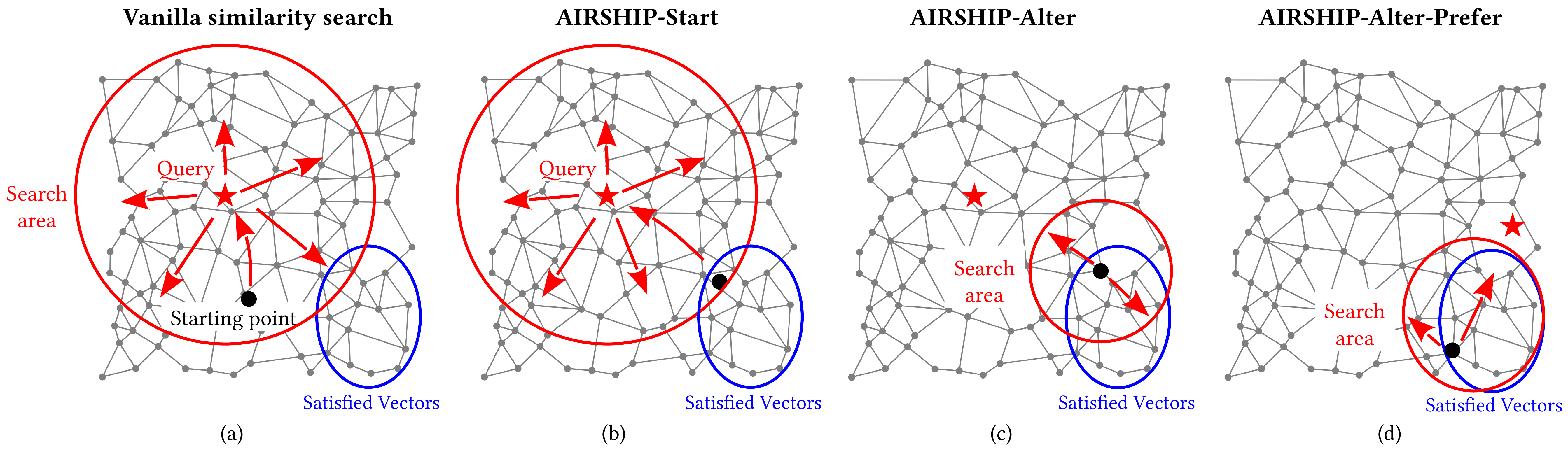}
\caption{Examples for the vanilla similarity search, AIRSHIP-Alter, and AIRSHIP-Alter-Prefer.}
\label{fig:airship-diff}
\end{figure*}

Note that this vanilla algorithm is inefficient when the query vector does not locate near the satisfied vectors. Figure~\ref{fig:airship-diff}(a) presents an example: The searching algorithm goes from the starting point toward the query vector, however, no satisfied vectors can be found in that path. The searching algorithm keeps expanding the searching radius and cannot touch any satisfied vector until it enumerates a lot of very similar but unsatisfied vectors.

\subsection{Starting Points Selection}
Our first optimization in the proposed AIRSHIP system is to locate good starting points for the graph searching algorithm. We denote this method as \texttt{AIRSHIP-Start}. We would like to start the graph search inside an ``interesting region''---a cluster of satisfied vectors. In this scenario (Figure~\ref{fig:airship-diff}(b)), a small number of satisfied vectors can be retrieved along the path to the query vector.

\vspace{0.1in}
\noindent
\textbf{Linear Scan.}
The most straightforward solution to find satisfied vectors is linear scan that goes over all vectors and checks whether it meets the constraints. Consider the time complexity of evaluating the constraint (a user-defined function) is $O(f)$. The linear scan takes $O(nf)$ time, where $n$ is the number of base vectors. Note that a brute force nearest neighbor search requires $O(nd)$ to compute distances between the query vector and all query vectors. Common constraints (e.g., checking if the attribute of a vector belongs to a range or a collection of categories) takes constant time---$O(1)$. In this case, the time complexity of linear scan is $O(n)$---it is $d$ times faster than the brute force nearest neighbor search especially when we are tackling with high dimensional data (i.e. $d > 100$). We urge that this linear scan is necessary for selective query constraints---there are only a small/constant number of satisfied vectors. Because we do not know the query constraints in the pre-processing stage, it is infeasible to build an index for each possible constraint. In addition, even if we have prior knowledge of the constraints, it is expensive to construct and store all those indices. The indices roughly require $O(nc)$ time and space, where $c$ is the number of distinct constraints---$c$ can be as large as the number of the queries. In this extreme scenario, after retrieved the small constant number of satisfied vectors, we do not need any advanced techniques other than a brute force distance computation to rank these vectors. Thus, as mentioned above in Section~\ref{sec:intro}, we have the following assumption:
\begin{assumption}
For each constraint, at least $p\%$ of base vectors are satisfied vectors. Formally,
$$\left\vert\left\{v_{i}|f_{j}\left(v_{i}\right) \rightarrow \texttt{true}, v_{i} \in V\right\}\right\vert \geq \frac{p}{100} \cdot n \hspace{.5cm} \forall f_{j}$$
\label{assum:at_least}
\end{assumption}
This assumption holds for real-world recommendation system applications, e.g., the query constraints filter out the retrieved vectors from a collection of categories.
When Assumption~\ref{assum:at_least} does not hold, a linear scan can be employed to retrieve the satisfied vectors and brute force distance computation is applied to rank them.

\vspace{0.1in}
\noindent
\textbf{Sampling.}
Although the linear scan that checks the constraints is faster than a brute force similarity search, it still requires us an inevitable linear time to perform the similarity search. In
Here we introduce our second assumption on the data distribution:
\begin{assumption}
Satisfied vectors are clustered in a constant number ($r$) of regions---they are not randomly distributed.
\label{assum:cluster}
\end{assumption}
This assumption commonly holds in the real-world datasets when the constraints are applied to the data labels---vectors with the same label are clustered in the high-dimensional (embedded) space. On the other hand, when the assumption does not hold, consider the case that the satisfied vectors of a given query constraint are randomly distributed, i.e., a vector satisfies the constraint with a fixed probability \textit{prob}.
Given this kind of constraints, when we target to find top-$k$ candidates, it is equivalent to retrieve top-$(\frac{k}{\textit{prob}})$ results in expectation with any similarity search techniques---we can view the fixed satisfied probability \textit{prob} as the probability to keep a retrieved item in the top candidates. Therefore, no optimization can be applied to the randomly distributed satisfied vectors.

With Assumption~\ref{assum:cluster}, we do not have to retrieve all satisfied vectors as the linear scan method, because satisfied vectors are clustered as regions and we can perform similarity search within these regions. Therefore, we are only required to locate a sample of satisfied vectors as searching starting points for the proximity graph searching algorithm. We obtain a sample of $s$ base vectors when we construct the graph index in the pre-processing stage. Due to Assumption~\ref{assum:at_least}, we have $p\cdot s$ satisfied vectors in the sample in expectation. The starting point selection in the sample takes $O(s)$ time for constant-time constraints.

\vspace{0.1in}
\noindent
\textbf{Initialization with multiple starting points.}
After we obtain a collection of satisfied vectors from the sampling method, a straightforward question is: which vector should we use as the starting point? We put all of them into the priority queue. The priority queue extracts the vector that is closest to the query point at the beginning of the search. Instead of simply using the single closest satisfied vector as the starting point, our approach enables other vectors can still be extracted during the search---when the previously extracted vectors connect to the candidates that are far away from the query point.

\vspace{0.1in}
\noindent
\subsection{Multi-Direction Search}
Although \texttt{AIRSHIP-start} alleviates the drawback of the vanilla solution, it is still inefficient when we want to retrieve a larger number of similar vectors (a larger $K$). Figure~\ref{fig:airship-diff} shows that only the satisfied vectors along the path to the query vector are retrieved at the beginning of the search. When we try to obtain more satisfied vectors, we still have to first navigate to the query point and then enumerate numerous similar but unsatisfied vectors.

\begin{algorithm}[htbp]
\algsetup{linenodelimiter=.}
\caption{AIRSHIP-Alter}\label{alg:airship-alter}
\begin{flushleft}
\textbf{Input:} A Proximity graph $G(V,E)$; a collection of starting points $S$; a query vector $q$; a user-defined query constraint $f$; a number of output candidates $K$.\\
\textbf{Output:} Top-$K$ vectors for the query $q$ with the constraint $f$.
\end{flushleft}
\begin{spacing}{0.9}
\begin{algorithmic}[1]
\STATE Initialize two min-heap priority queues $\textit{pq}_{\textit{sat}}$ and $\textit{pq}_{\textit{other}}$; a max-heap priority queue $\textit{topk}$.
\STATE $\textit{visited} \leftarrow \emptyset$
\FOR{\textbf{each} $S_i \in S$}
    \STATE $\textit{d} \leftarrow \textit{dist}(S_i,q)$
    \STATE $\textit{pq}_{\textit{sat}} \leftarrow \textit{pq}_{\textit{sat}} \cup \{(d,S_i)\}$
    \STATE $\textit{visited} \leftarrow \textit{visited} \cup \{S_{i}\}$
\ENDFOR
\WHILE{$\textit{pq}_{\textit{sat}} \not= \emptyset$ \textbf{and} $\textit{pq}_{\textit{other}} \not= \emptyset$}
    \STATE $\textit{pq} \leftarrow {\small \textit{select\_priority\_queue}}(\textit{pq}_{\textit{sat}},\textit{pq}_{\textit{other}},\textit{cnt}_{\textit{sat}},\textit{cnt}_{\textit{total}},\textit{alter\_ratio})$\label{line:select-pq}
    \IF{$\textit{pq} = \textit{pq}_{\textit{sat}}$}\label{line:maintain-cnt-begin}
        \STATE $\textit{cnt}_{\textit{sat}} \leftarrow \textit{cnt}_{\textit{sat}} + 1$
    \ENDIF
    \STATE $\textit{cnt}_{\textit{total}} \leftarrow \textit{cnt}_{\textit{total}} + 1$\label{line:maintain-cnt-end}
    \STATE $(\textit{now\_dist},\textit{now\_idx}) \leftarrow \textit{pq}.\textit{pop\_min}()$
    \IF{$|\textit{topk}| = K$ \textbf{and} $\textit{now\_dist} > \textit{topk}.\textit{peek\_max}()$}
        \STATE \textbf{break}
    \ENDIF
    \IF{$\textit{pq} = \textit{pq}_{\textit{sat}}$}\label{line:update-begin}
        \STATE $\textit{topk} \leftarrow \textit{topk} \cup \{\textit{now\_dist},\textit{now\_idx}\}$
        \IF{$|\textit{topk}| > K$}
            \STATE $\textit{topk}$.\textit{pop\_max}()
        \ENDIF
    \ENDIF\label{line:update-end}
    \FOR{\textbf{each} \textit{next\_idx} $\in$ neighbors of \textit{now\_idx} in $G$}
        \IF{$\textit{next\_idx} \not\in \textit{visited}$}
            \STATE $\textit{d} \leftarrow \textit{dist}(\textit{next\_idx},q)$
            \IF{$f(next\_idx) = \textbf{true}$}\label{line:two-queue-insert-begin}
                \STATE $\textit{pq}_{\textit{sat}} \leftarrow \textit{pq}_{\textit{sat}} \cup \{ (d,\textit{next\_idx})\}$
            \ELSE
                \STATE $\textit{pq}_{\textit{other}} \leftarrow \textit{pq}_{\textit{other}} \cup \{ (d,\textit{next\_idx})\}$
            \ENDIF\label{line:two-queue-insert-end}
            \STATE $\textit{visited} \leftarrow \textit{visited} \cup \{\textit{next\_idx}\}$
        \ENDIF
    \ENDFOR
\ENDWHILE
\RETURN \textit{topk}
\end{algorithmic}
\end{spacing}
\end{algorithm}

\vspace{0.1in}
\noindent
\textbf{AIRSHIP-Alter.} Following this intuition, we propose an optimization AIRSHIP-Alter that enables multi-direction searching. Comparing with the single priority queue used in the original graph search, we have two priority queues to keep search candidates (Algorithm~\ref{alg:airship-alter}). $\textit{pq}_{\textit{sat}}$ stores the satisfied vectors and $\textit{pq}_{\textit{other}}$ maintains remaining unsatisfied vectors. Line~\ref{line:select-pq} selects a priority queue by Algorithm~\ref{alg:select-pq}. When both priority queues are not empty, we choose each queue with a weight ratio---\textit{alter\_ratio} (Algorithm~\ref{alg:select-pq} Line~\ref{line:ratio-select}). The \textit{alter\_ratio} ($0 < \textit{alter\_ratio} \leq 1$) is used to determine the proportion of search candidates from the $\textit{pq}_{\textit{sat}}$.  For example, both queues are selected alternatively after the other when $\textit{alter\_ratio} = 0.5$. The greater the \textit{alter\_ratio} is, the more biased we are to select the priority queue with satisfied vectors ($\textit{pq}_{\textit{sat}}$).
Back to Algorithm~\ref{alg:airship-alter}, Line~\ref{line:maintain-cnt-begin}-\ref{line:maintain-cnt-end} maintain the number of times we select $\textit{pq}_{\textit{sat}}$ ($\textit{cnt}_{\textit{sat}}$) and count the total number of processed iterations ($\textit{cnt}_{\textit{total}}$). Once the priority queue is determined, we extract the search candidate from the queue (Line~\ref{line:update-begin}-\ref{line:update-end}) as the original searching algorithm. Another major difference from the original searching algorithm is Line~\ref{line:two-queue-insert-begin}-\ref{line:two-queue-insert-end}: We insert the neighbors of the current searching vector into two separate priority queues according to their constraint satisfactory---satisfied vectors are assigned to $\textit{pq}_{\textit{sat}}$ and the remaining ones are pushed to $\textit{pq}_{\textit{other}}$.

Searching with the candidates from $\textit{pq}_{\textit{sat}}$ ``exploits'' the clusterness of satisfied vectors, while using the candidates from $\textit{pq}_{\textit{other}}$ ``explores'' other regions that close to the query vector. This two-priority-queue design enables us to bias satisfied vectors and have a trade-off between the ``exploitation'' and ``exploration''. Figure~\ref{fig:airship-diff}(c) presents an example when $alter\_ratio=0.5$, we alternatively extract the vectors from two priority queues and a good amount of satisfied vectors are retrieved.

\clearpage

\subsection{alter\_ratio Estimation}
As a hyper-parameter, \textit{alter\_ratio} can be enumerated and tuned for applications to achieve the best performance. It is also common to have no prior knowledge of the query constraints,  i.e., no pre-tuning can be applied. In this section, we derive a low-overhead heuristic algorithm to estimate \textit{alter\_ratio}. Intuitively, \textit{alter\_ratio} should be positively correlated with the clusterness of satisfied vectors: The more clustered satisfied vectors, the larger \textit{alter\_ratio} should be. In the extreme case when all satisfied vectors are distributed in the same region and no unsatisfied vectors are within this region, \textit{alter\_ratio} should be $1$ to obtain the optimal performance, as we do not need to explore unsatisfied vectors. Here, we adapt the concept of k-nearest-neighbor statistics~\citep{hastie2009elements} as the estimation of \textit{alter\_ratio}. With a given constraint $f$ and a collection of satisfied vectors from the samples ($SSV$), our estimation of \textit{alter\_ratio} is:
\begin{equation}
\label{eq:alter-ratio}
\frac{1}{|SSV|}\sum_{v \in SSV}{ \frac{|{\{v' | f(v')=\texttt{true} \hspace{.1cm} \forall v' \in \textit{top-k\_neighbor}(v)\}}|}{k}   }
\end{equation}
where $\textit{top-k\_neighbor}(v)$ is the nearest $k$ neighbors of vertex $v$. Note that we do not need any extra computation to obtain these neighbors in the query time. Since the proximity graph approximates a k-nearest-neighbor graph by its definition, we can view the immediate neighbors of $v$ in the proximity graph as its nearest $k$ neighbors. We sort the edges in the proximity graph according to their distance to the vertex. Therefore, no distance computation is required in the query time to obtain the estimation of \textit{alter\_ratio}. For each sampled satisfied vectors, we only need to go over its first $k$ edges and check whether its neighbors are satisfied vectors. \textit{alter\_ratio} is estimated as the average number of satisfied neighbors among all sampled satisfied vectors.

\begin{algorithm}[b!]
\algsetup{linenodelimiter=.}
\caption{select\_priority\_queue}\label{alg:select-pq}
\begin{flushleft}
\textbf{Input:} Two priority queues $\textit{pq}_{\textit{sat}}$ and $\textit{pq}_{\textit{other}}$; $\textit{cnt}_{\textit{sat}}$; $\textit{cnt}_{\textit{total}}$; \textit{alter\_ratio}.\\
\textbf{Output:} A priority queue for the current searching iteration.
\end{flushleft}
\begin{spacing}{0.9}
\begin{algorithmic}[1]
    \IF{$(\textit{pq}_{\textit{other}} = \emptyset)$}
        \STATE $\textit{pq} \leftarrow \textit{pq}_{\textit{sat}}$
    \ELSIF{$\textit{pq}_{\textit{sat}} = \emptyset$}
        \STATE $\textit{pq} \leftarrow \textit{pq}_{\textit{other}}$
    \ELSIF{$\frac{\textit{cnt}_{\textit{sat}}}{\textit{cnt}_\textit{total}} \leq \textit{alter\_ratio}$}
        \STATE $\textit{pq} \leftarrow \textit{pq}_{\textit{sat}}$\label{line:ratio-select}
    \ELSE
        \STATE $\textit{pq} \leftarrow \textit{pq}_{\textit{other}}$
    \ENDIF
    \RETURN \textit{pq}
\end{algorithmic}
\end{spacing}
\end{algorithm}

\subsection{Biased Priority Queue Selection}\label{ssec:prefer}
As depicted in Figure~\ref{fig:airship-diff}(d), when the path from the starting point to the query vector are inside the region of satisfied vectors, the exploration of $\textit{pq}_{\textit{other}}$ are doing ``useless'' efforts, as it is not exploring the direction toward the query vector. These ineffective operations are brought by  parameter \textit{alter\_ratio} that forces us to explore unsatisfied vectors. We propose \texttt{AIRSHIP-Alter-Prefer} to overcome this disadvantage: when the top candidate from the satisfied vector queue is better than the one from the other queue, we override the $alter\_ratio$ restriction and select the satisfied vector queue. This optimization is realized by adding another \texttt{if} statement in the priority queue selection (Algorithm~\ref{alg:select-pq}): if the top candidate from $\textit{pq}_{\textit{sat}}$ is better than the one from $\textit{pq}_{\textit{other}}$, we select $\textit{pq}_{\textit{sat}}$ without considering \textit{alter\_ratio}.
This optimization is aggressive: it benefits the searching when the satisfied vectors are highly clustered, while it also reduces the explorations of unsatisfied vectors---what may hurt the performance when top similar satisfied vectors do not reside near the clusters of sampled starting points. We investigate this optimization in Section~\ref{ssec:exp-prefer}.

\section{Experimental Evaluation}
The objective of the experimental evaluation is to investigate the overall performance and the impact of optimizations of the proposed system \texttt{AIRSHIP}. Specifically, the experiments are designed to answer the following questions:
\begin{itemize}
\item How does the proposed system compare with other state-of-the-art similarity search algorithms?
\item What is the effect of the proposed optimizations, namely AIRSHIP-Start, AIRSHIP-Alter, and AIRSHIP-Alter-Prefer?
\item How is our estimation to \textit{alter\_ratio} on different data distributions? Does searching performance with the estimated ratio match the best enumerated/hand-picked constant?
\item How does the data distribution affect the performance of the proposed system?
\end{itemize}

\vspace{0.1in}
\noindent
\textbf{Implementation.}
We implement the vanilla graph-based similarity search and AIRSHIP as a C++11 prototype. The code is compiled with g++-7.5.0 enabling the ``O3'' optimization. The code contains special function calls to harness detailed profiling data. 

\vspace{0.1in}
\noindent
\textbf{System.}
We execute the experiments on a single
node server. The server has one Intel(R) Xeon(R) Platinum 8276 CPU @ 2.20GHz
(64 bit)---28 cores 56 threads---and 1 TB of memory. Ubuntu 20.04 LTS 64-bit is the operating system.

\vspace{0.1in}
\noindent
\textbf{Baselines.}
We have two baseline algorithms in the experiments: \textbf{Vanilla} and \textbf{product quantization (PQ)}~\citep{jegou2011product}. Vanilla, which uses the same code base of~\citet{zhao2020song}, is an adaption of HNSW~\citep{malkov2020efficient} to our constrained similarity search problem. We add the query constraint filter before pushing top-k results into the result heap. All our proposed AIRSHIP optimizations are built on top of the vanilla solution. PQ is one of the most popular quantization-based similarity search algorithms. It quantizes the high-dimensional vector to a limited number of integers to accelerate the distance computation. For the PQ comparison, we go over all vectors and check whether they meet the query constraint. Then, top-k similar vectors are chosen by the distances on the quantized vectors instead of the original data vector.

\vspace{0.1in}
\noindent
\textbf{Methodology.}
We pre-build the index for each algorithm as a pre-processing---the index construction time is not included in the experiments. All experiments are executed with one single thread to have a fair comparison. Since all algorithms support inter-query parallel, we can assume their performances are linearly scalable to the number of threads. We use Euclidean distance as the similarity measure---0 distance represents the highest similarity. The performance of an algorithm is evaluated by two factors: searching time (\textit{Query Per Second}) and retrieval quality (\textit{recall}).


\newpage

\noindent
\textbf{Searching Time}. We measure wall-clock time of each algorithm and present the number of completed Query Per Second (throughput) as the searching time measurement. We employ Query Per Second as the metric instead of the execution time of a query batch, because Query Per Second can be compared without normalizing query batch to the same size.
All experiments are performed 3 times and we report the average value as the result.

\vspace{0.1in}
\noindent
\textbf{Retrieval Quality.} Recall is a commonly-used retrieval quality measurement for similarity search algorithms. Suppose the candidate point set returned by an algorithm is $A$, and the correct $K$ nearest neighbors of the query is $B$, recall is defined as: $Recall(A) = \frac{|A \cap B|}{|B|}$.
A higher recall corresponds to a better approximation to the correct nearest neighbor result.

\vspace{0.1in}
\noindent
\textbf{Data.}
Two datasets are used in the experimental evaluation, one un-labeled data with synthetic labels and one real data with real labels. SIFT1m\footnote{\url{http://corpus-texmex.irisa.fr/}} has 1 million 128-dimensional vectors and does not have labels for each data vector. We cluster the vectors using k-means~\citep{jain2010data} ($k=10$) into $10$ clusters and assign the corresponding cluster id as the label for each vector. The randomness of the label is also studied. Consider $R\%$ randomness, when we assign the label for each vector, we have a probability $R\%$ to set a random label in $\{1,2,\cdots,10\}$ instead of its corresponding cluster id. We have $0\%$, $50\%$, and $100\%$ random labels on SIFT in the following experiments. The dataset with real labels we used is MNIST with 784 dimensions\footnote{\url{https://www.csie.ntu.edu.tw/~cjlin/libsvmtools/datasets/multiclass.html\#mnist8m}}. We use the first 1 million vectors in this paper to have the same base vector size as the SIFT dataset.

\vspace{0.1in}
\noindent
\textbf{Queries.}
The queries in the original similarity search datasets do not have constraints. We synthesize two categories of constraints. (a) \texttt{equal}: the returned similar vectors should have the same label as the query vector; (b) \texttt{unequal-X\%}: for each query, we randomly choose X\% labels that are unequal to the label of the query vector. The labels of returned similar vectors should be one of these randomly chosen labels. We extract 10,000 vectors from the testing data as the query vectors. The label of the SIFT1m query vector is generated in the same fashion as the base vectors.

\begin{figure*}[t!]
\begin{center}
 \mbox{\hspace{-0.1in}
    \includegraphics[scale = 0.39]{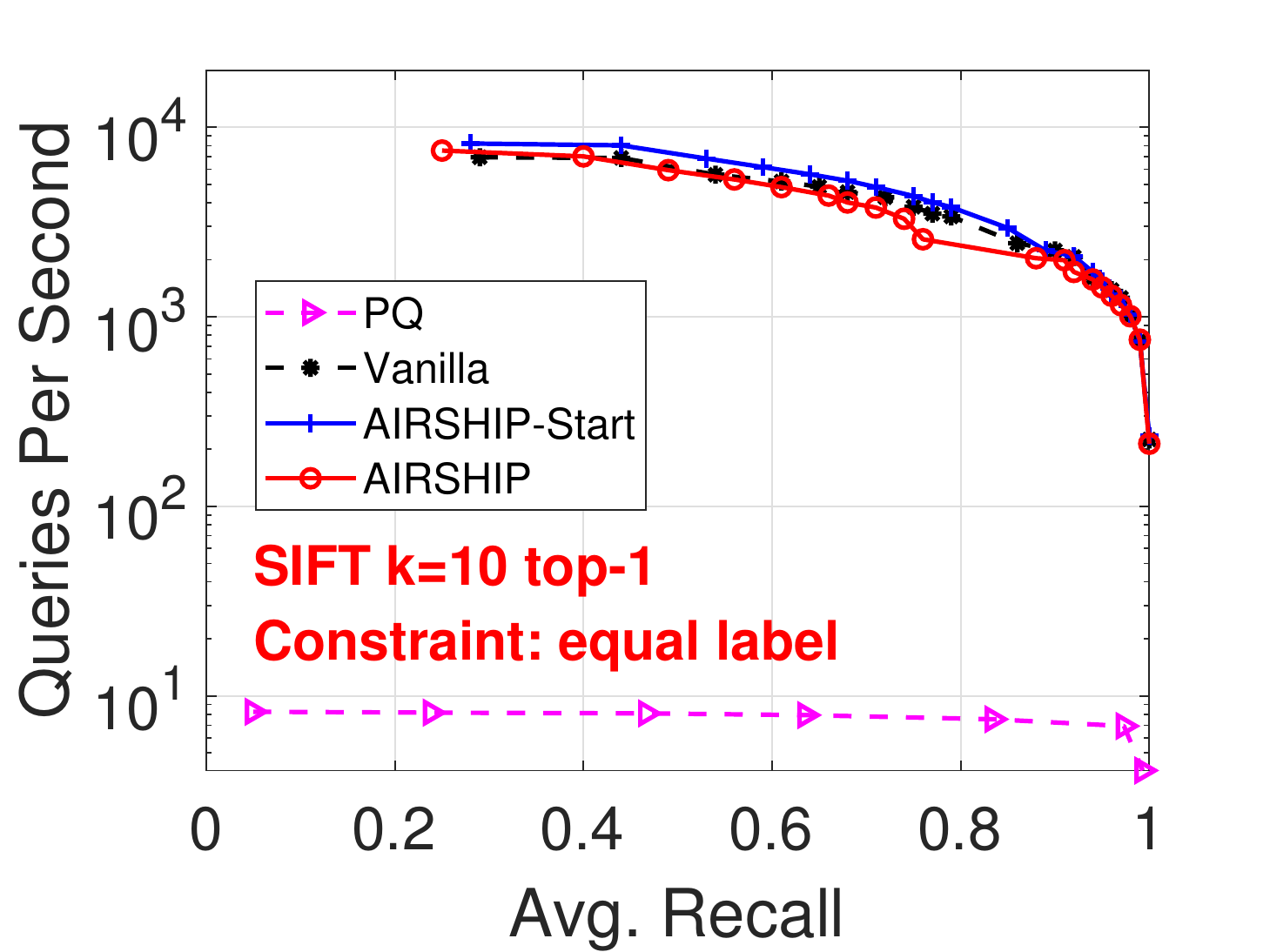}
    \hspace{-0.15in}
    \includegraphics[scale = 0.39]{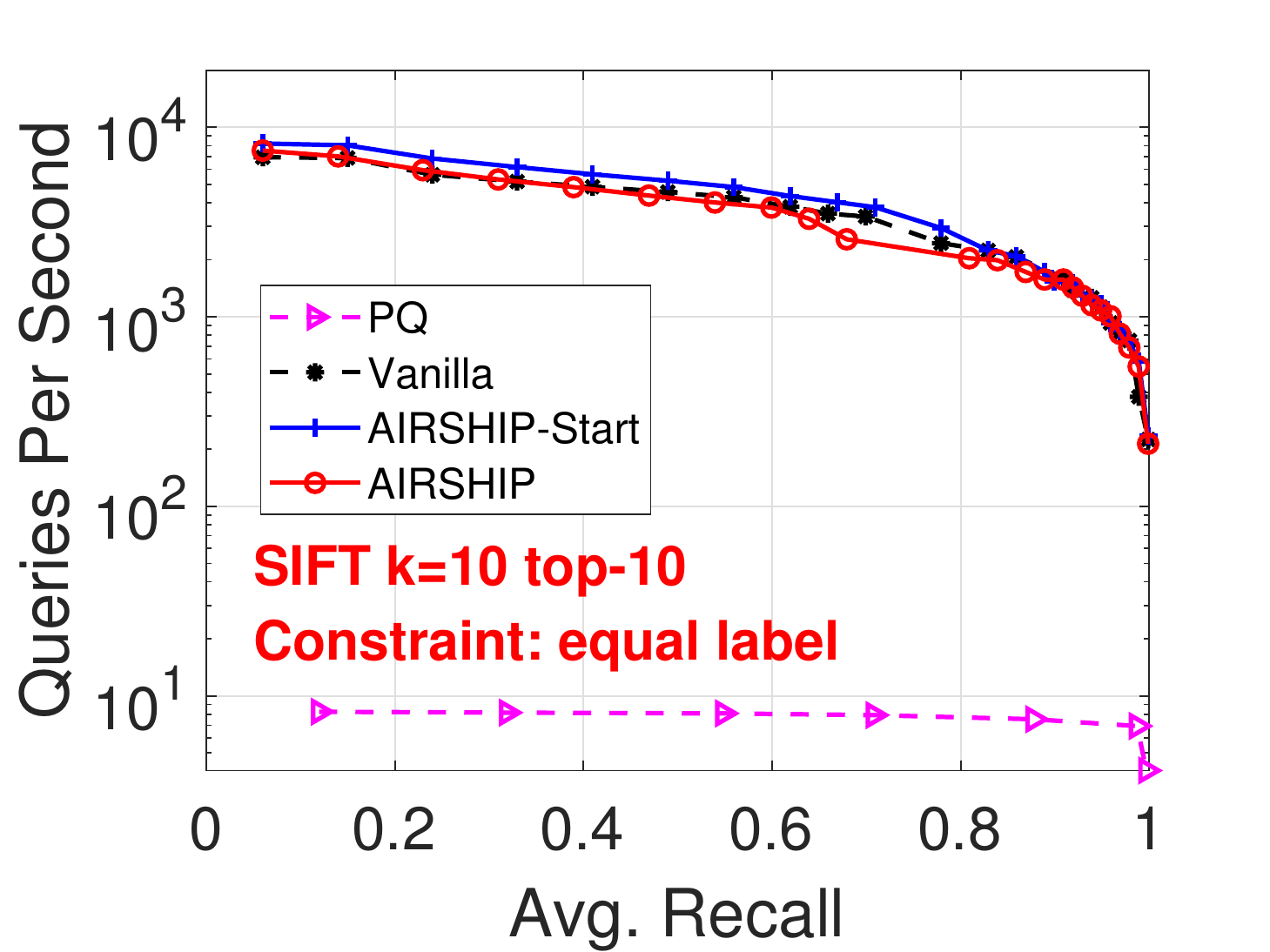}
    \hspace{-0.15in}
    \includegraphics[scale = 0.39]{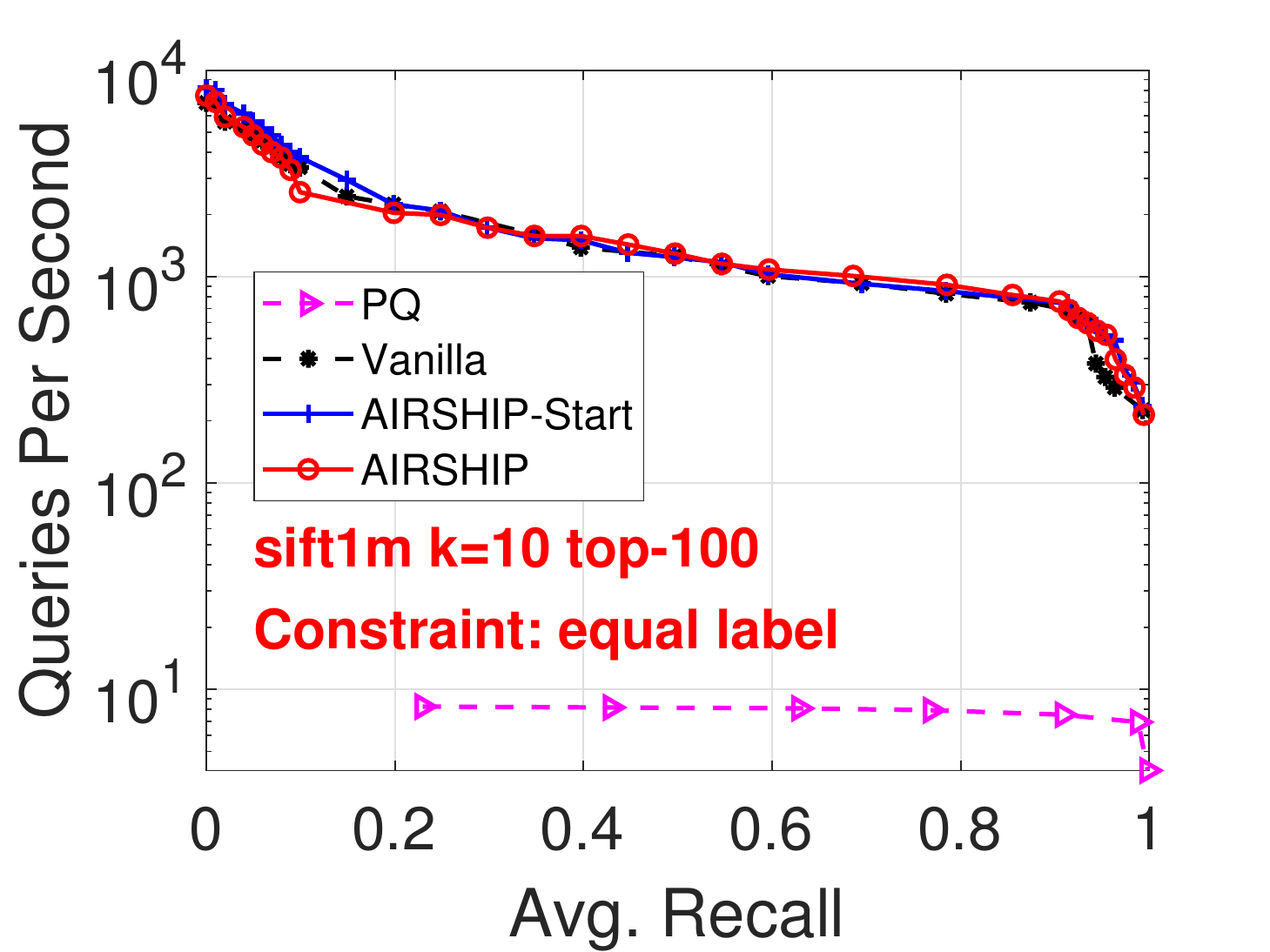}
    }
 \mbox{\hspace{-0.1in}
    \includegraphics[scale = 0.39]{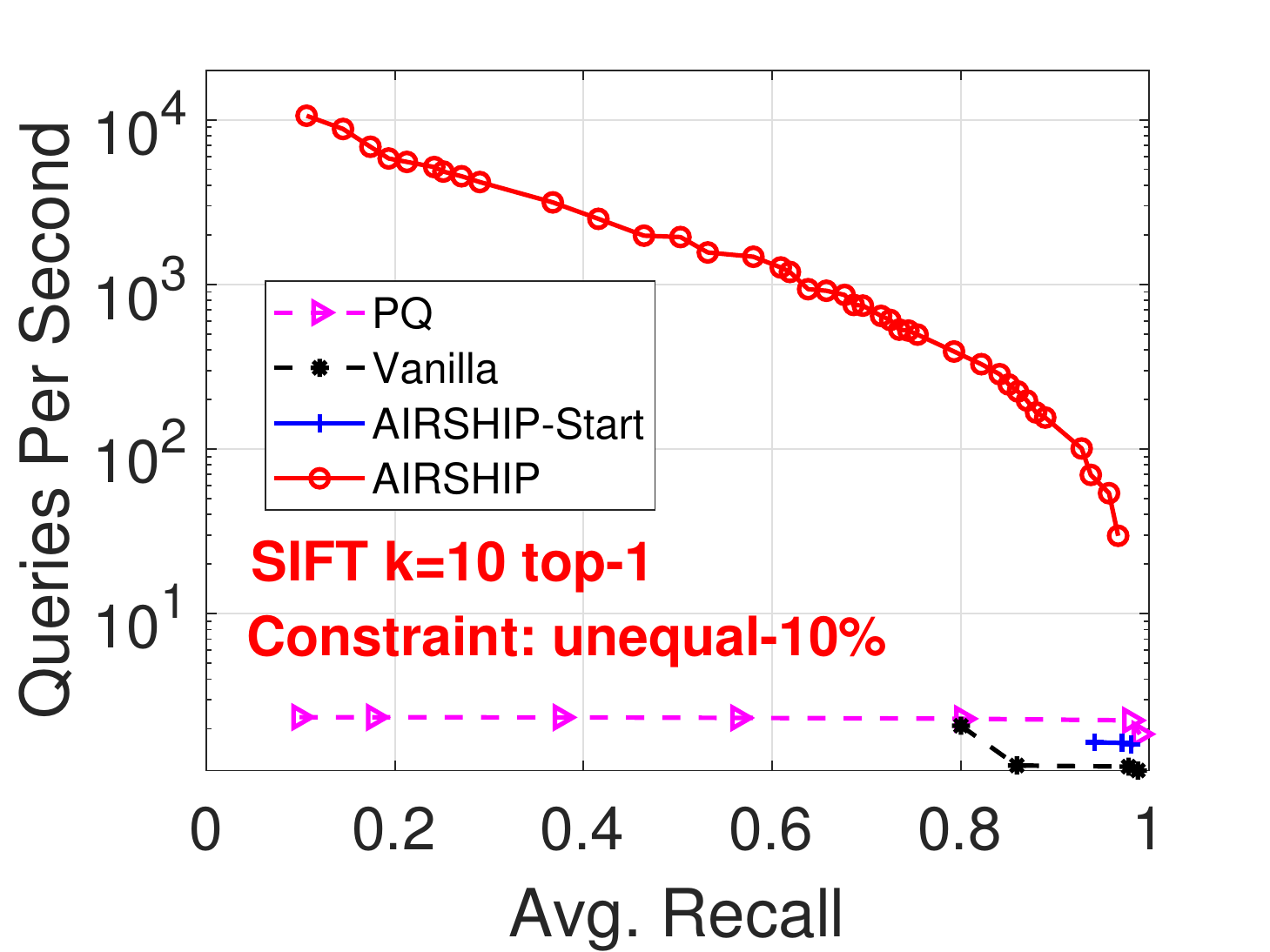}
    \hspace{-0.15in}
    \includegraphics[scale = 0.39]{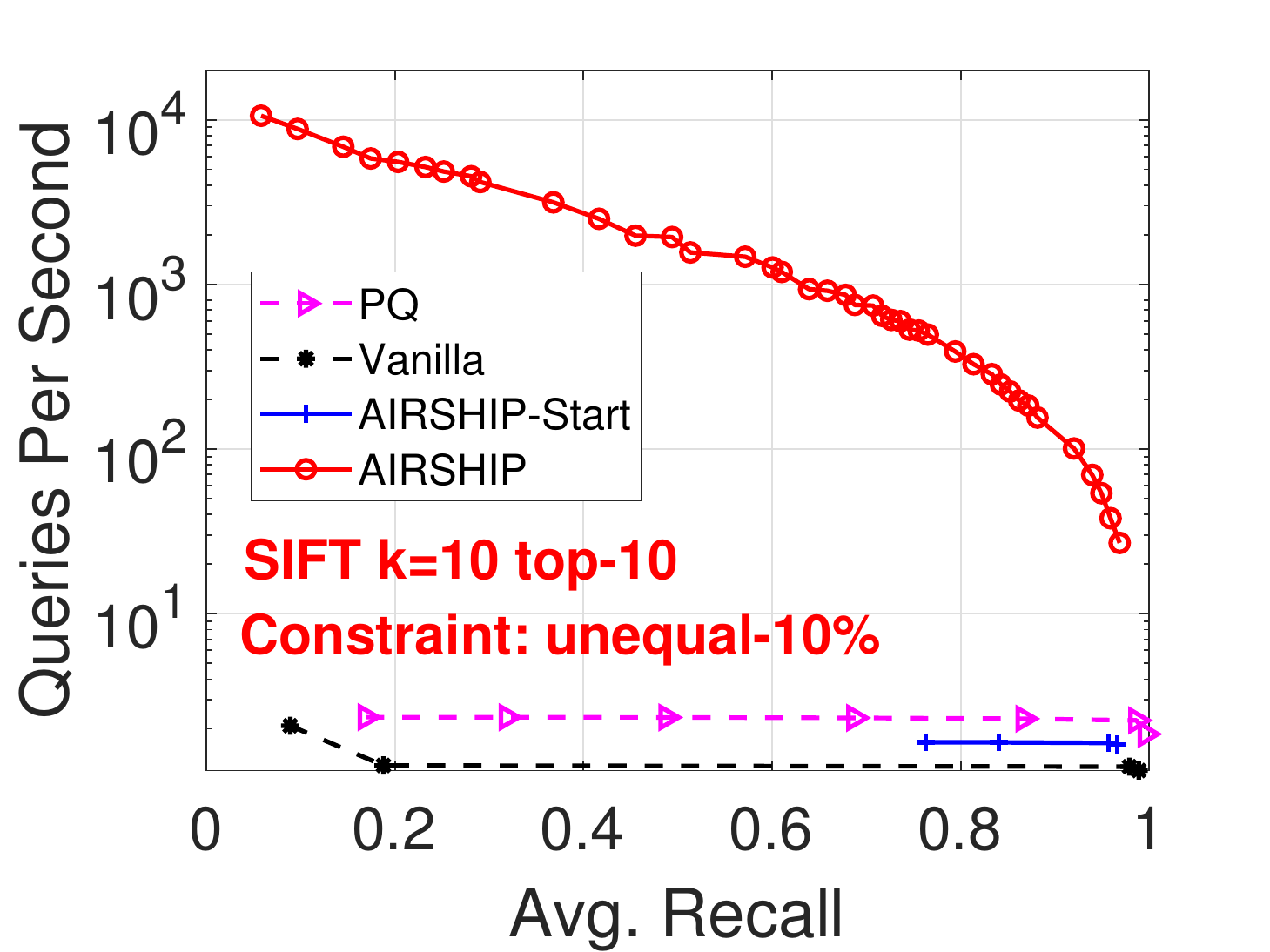}
    \hspace{-0.15in}
    \includegraphics[scale = 0.39]{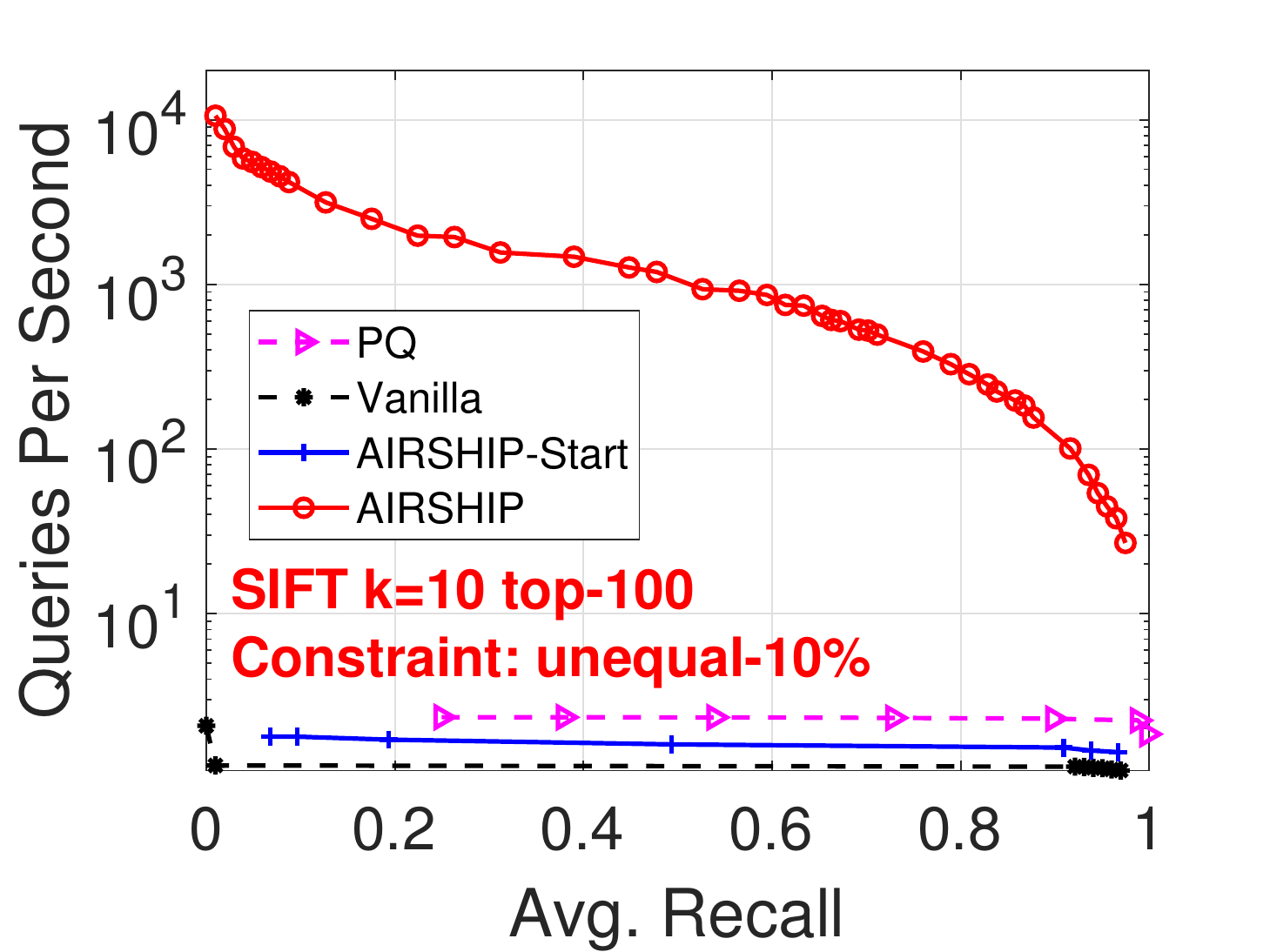}
    }
 \mbox{\hspace{-0.1in}
    \includegraphics[scale = 0.39]{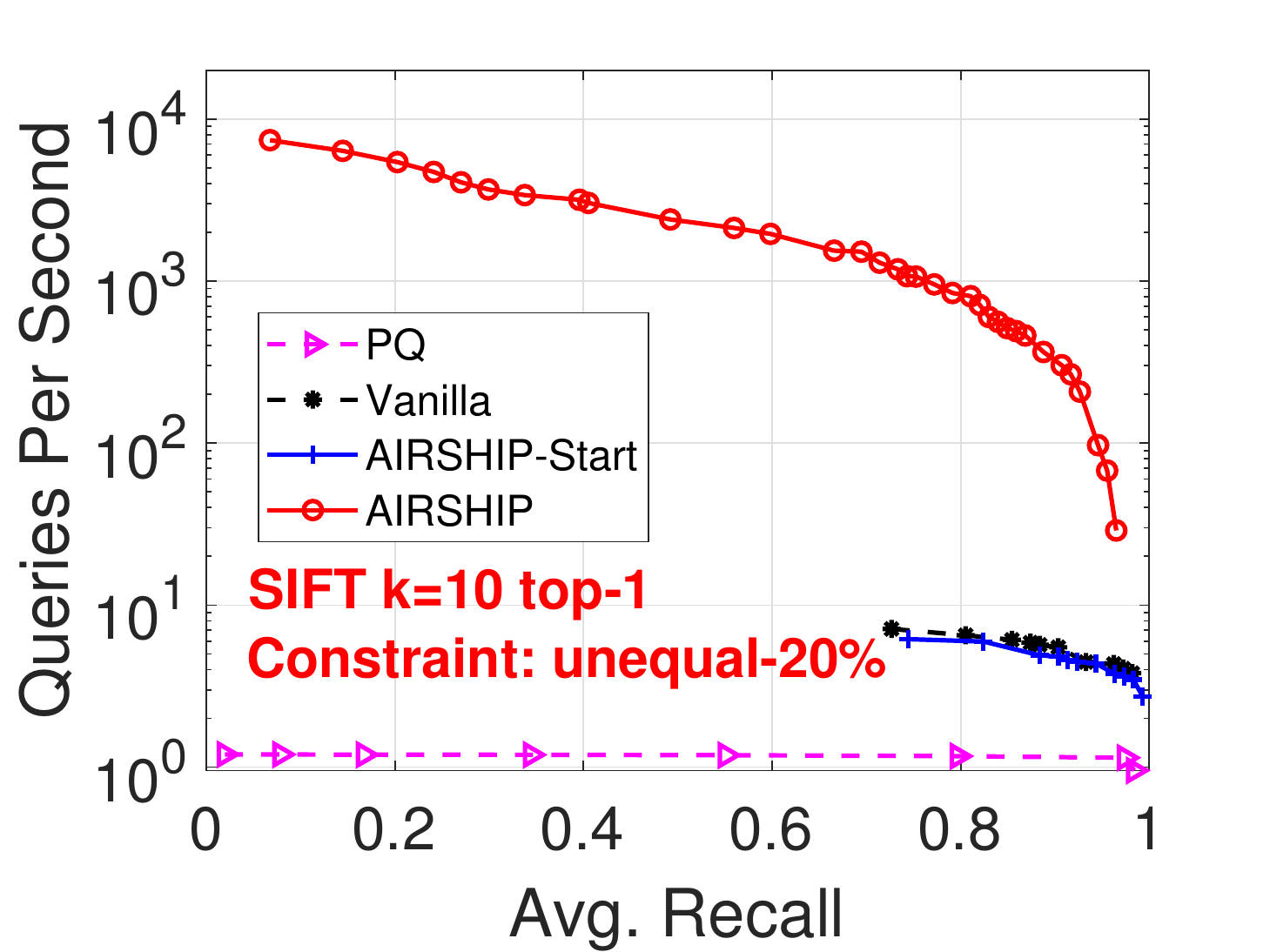}
    \hspace{-0.15in}
    \includegraphics[scale = 0.39]{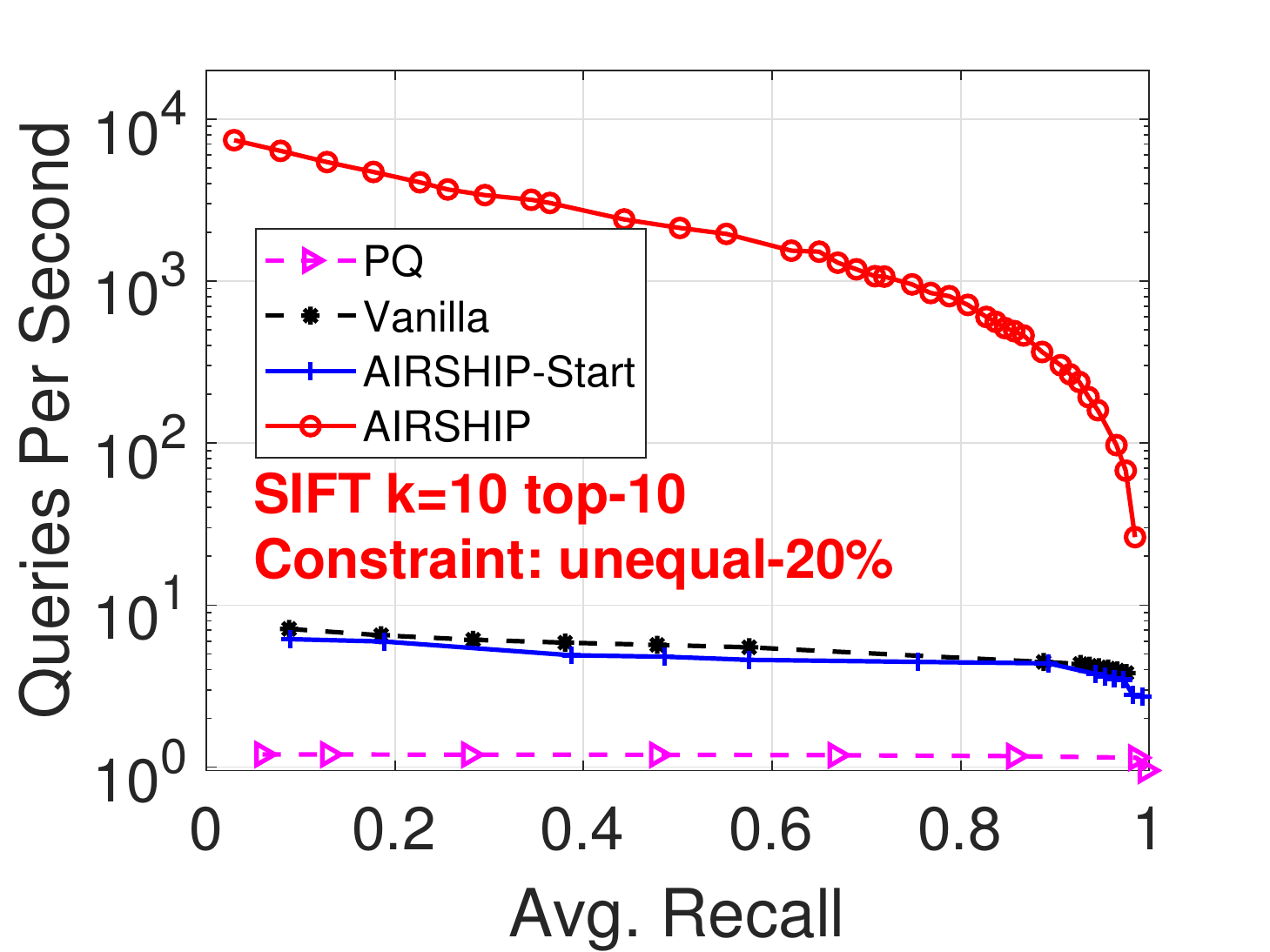}
    \hspace{-0.15in}
    \includegraphics[scale = 0.39]{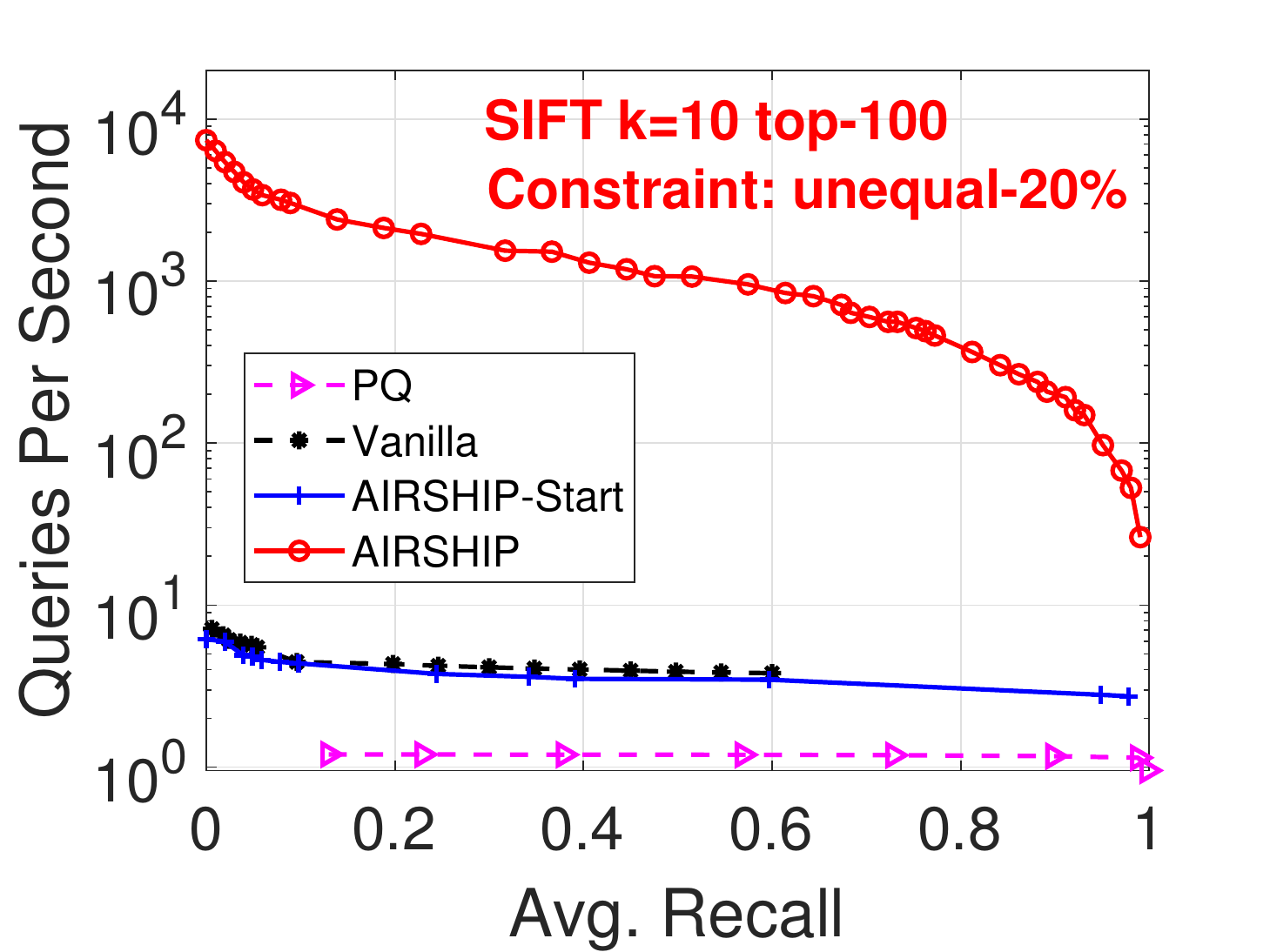}
    }
 \mbox{\hspace{-0.1in}
    \includegraphics[scale = 0.39]{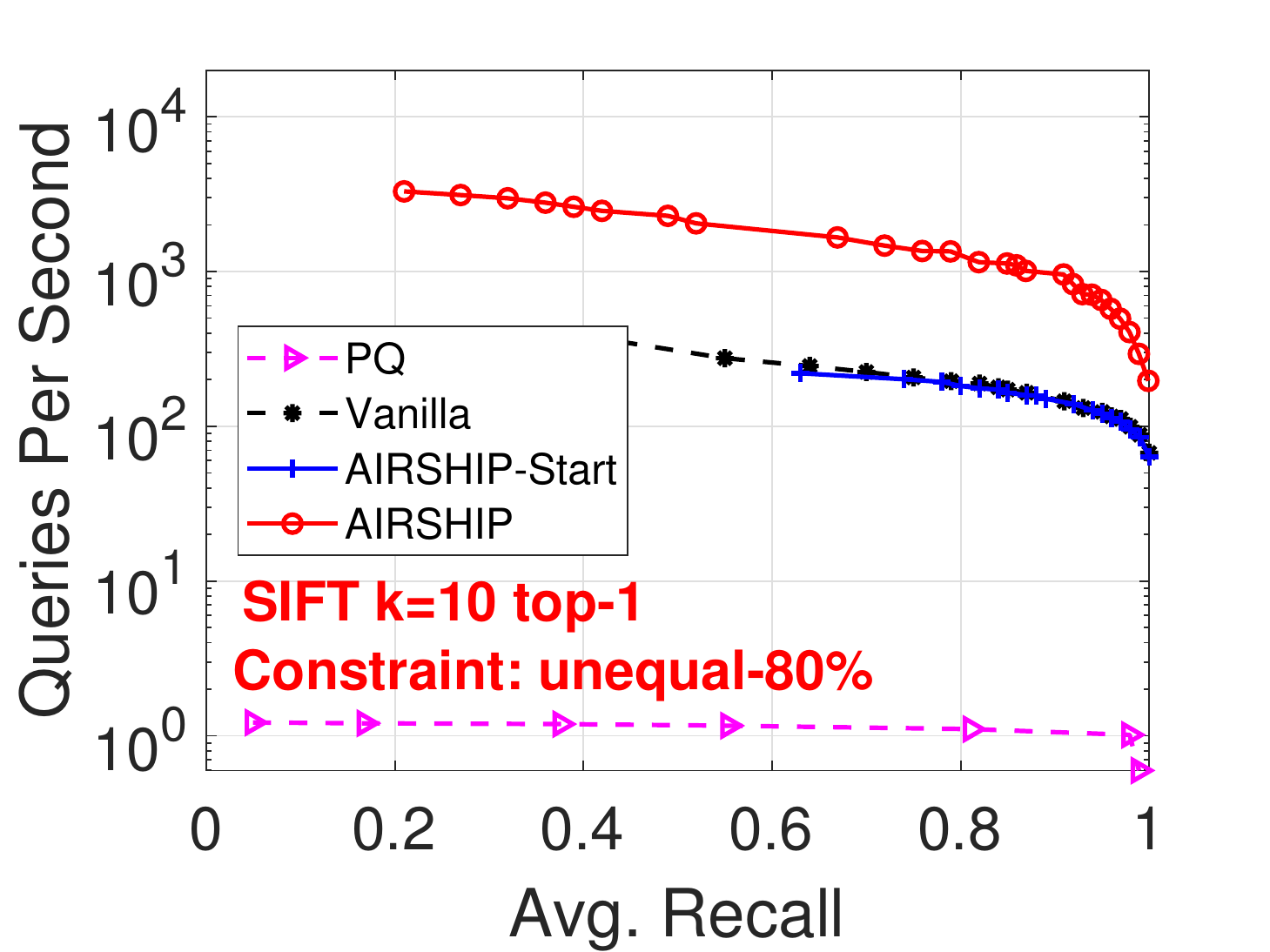}
    \hspace{-0.15in}
    \includegraphics[scale = 0.39]{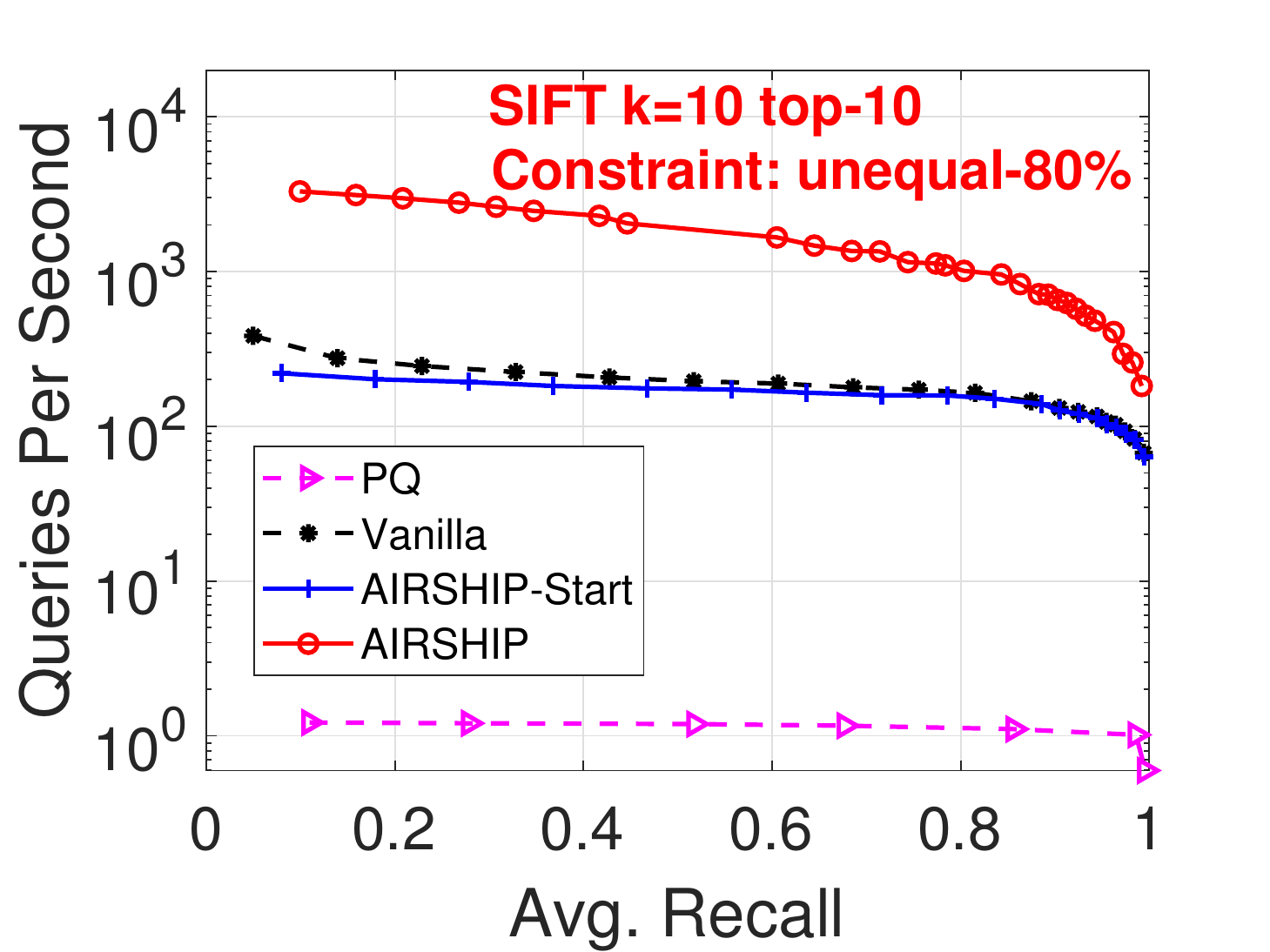}
    \hspace{-0.15in}
    \includegraphics[scale = 0.39]{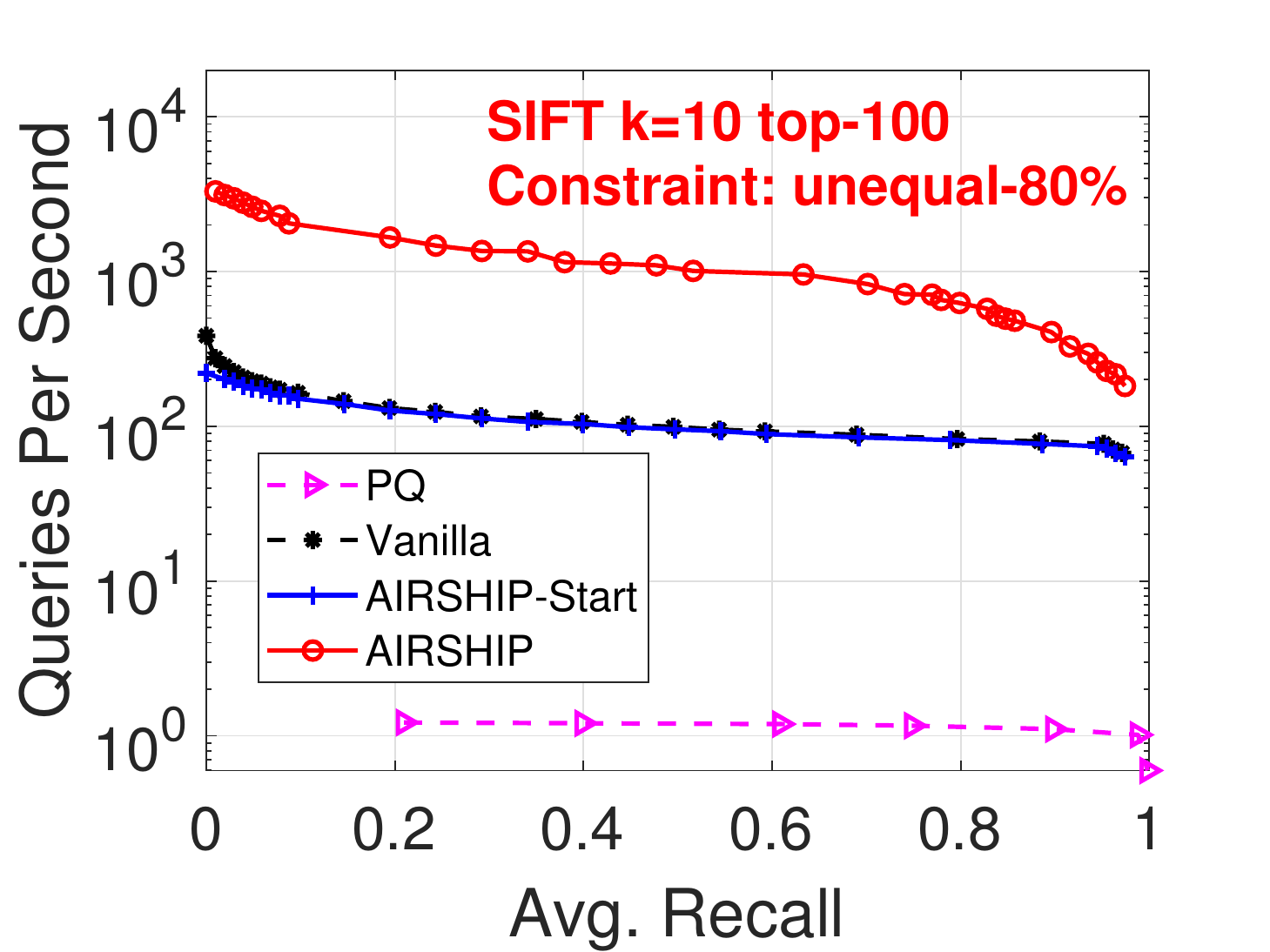}
    }

\end{center}
\vspace{-0.2in}
 \caption{Results on SIFT. The number of cluster $k=10$. Different constraints are applied.
}\label{fig:perf-rand0}  \vspace{0.1in}
\end{figure*}

\subsection{Performance Comparison}
Figure~\ref{fig:perf-rand0} depicts the comparison of PQ, vanilla graph-based similarity search, AIRSHIP-Start, and AIRSHIP---AIRSHIP is the algorithm with all proposed optimizations.

\vspace{0.1in}
\noindent
\textbf{Equal label queries.}
The first row of Figure~\ref{fig:perf-rand0} illustrate the comparison when our query constraints are: the returned similar vectors should have the same label as the query vector. Since the data with the same label is clustered---similar vectors have the same label to the query vector, this kind of constraint has no substantial difference to the original similar search without considering the constraints. Therefore, all 3 graph-based methods have comparable performance. PQ iterates all vectors to locate the satisfied vectors. Since it is linear to the number of base vectors, inferior performance is observed.

\vspace{0.1in}
\noindent
\textbf{Unequal label queries.}
When the query constraint targets to locate unequal labels---which is the setting of the constrained similarity search, substantial gaps are shown in the second to the fourth row of Figure~\ref{fig:perf-rand0}. Choosing a good starting point slightly improve the vanilla solution in \texttt{unequal-10\%}. For other constraints, vanilla and AIRSHIP-Start have almost identical performance. There is a substantial gap between AIRSHIP and other methods. Note that the y-axis is in the logarithmic scale. For \texttt{unequal-10\%}, AIRSHIP achieves a consistent 10-100x speedup for all top-1, 10, and 100 tasks. The gap is reduced with more relaxed constraints. In \texttt{unequal-20\%}, the gap becomes smaller because there are more satisfied vectors than \texttt{unequal-10\%}---the vanilla solution and AIRSHIP-Start have a greater chance to find the satisfied vectors. \texttt{unequal-80\%} is very similar to the unconstrained general similarity search problem. Thus, vanilla and AIRSHIP-Start shows better performance.

AIRSHIP has a consistent execution time across all $4$ query constraints. This confirms that our proposed solution is stable for various query constraints without building extra indices.

\begin{figure}
\begin{center}
 \mbox{
\includegraphics[width=2.45in]{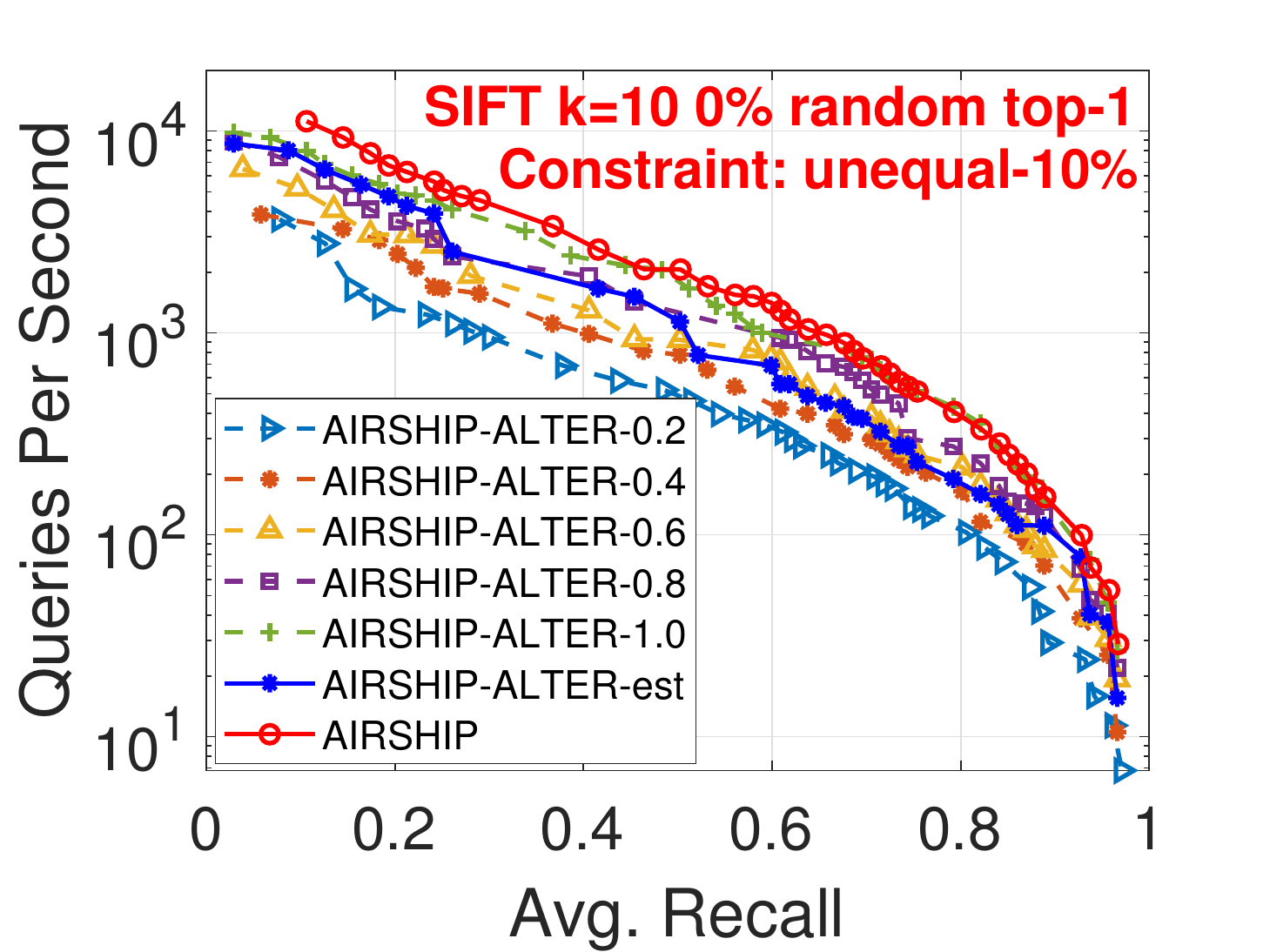}
\hspace{0.12in}
    \includegraphics[width=2.45in]{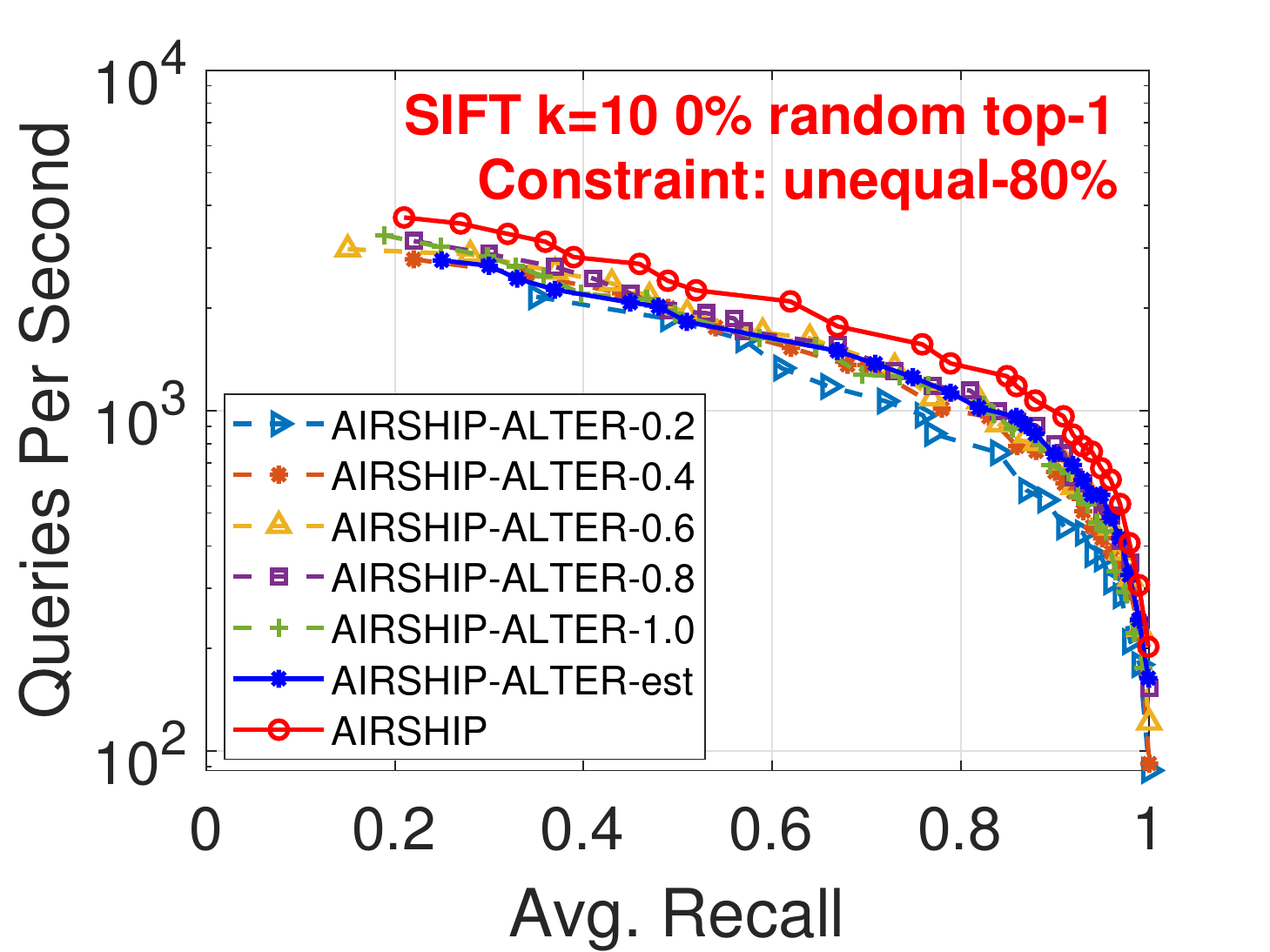}
}

\vspace{-0.14in}

\mbox{
\includegraphics[width=2.45in]{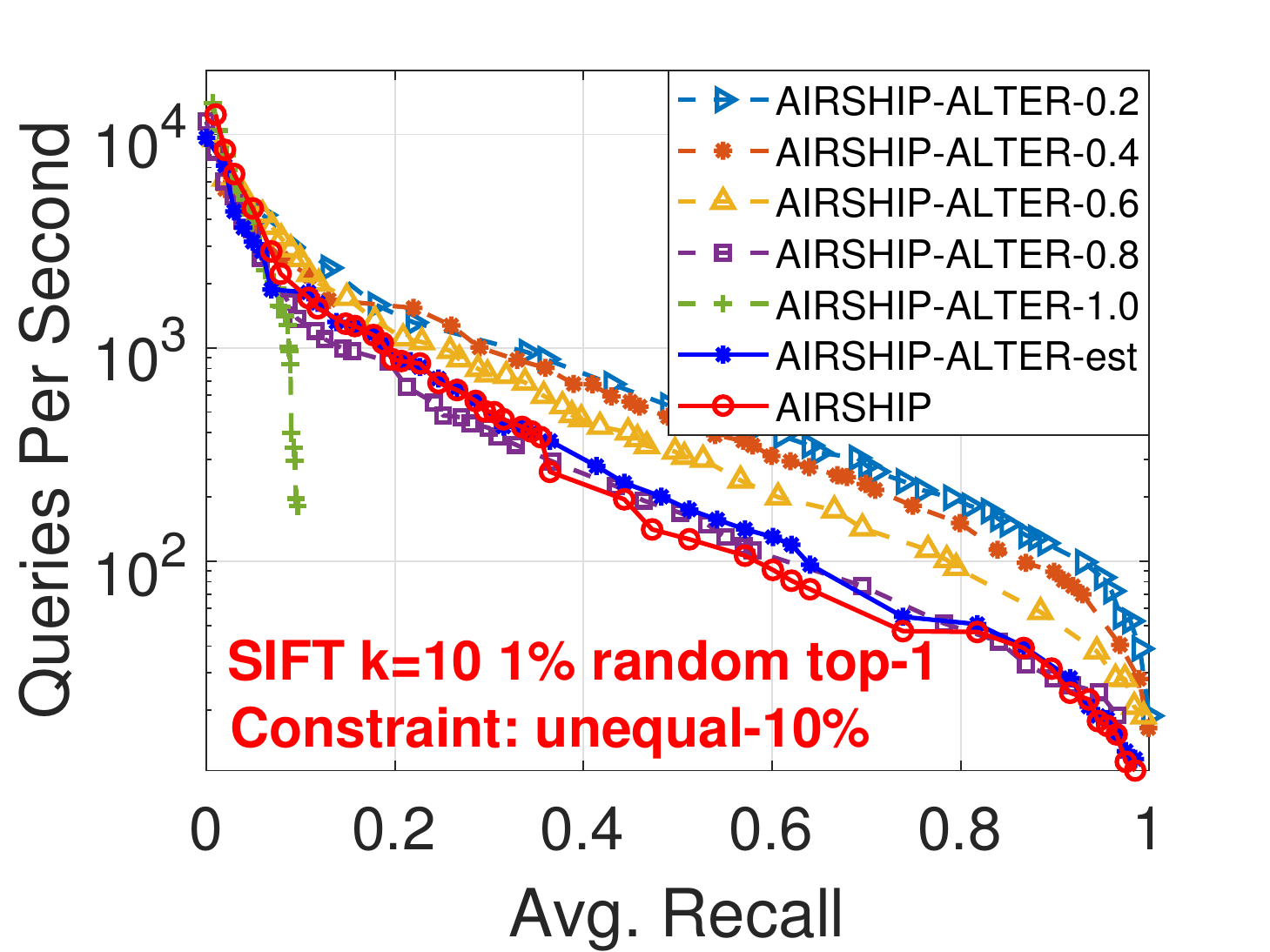}
\hspace{0.12in}
    \includegraphics[width=2.45in]{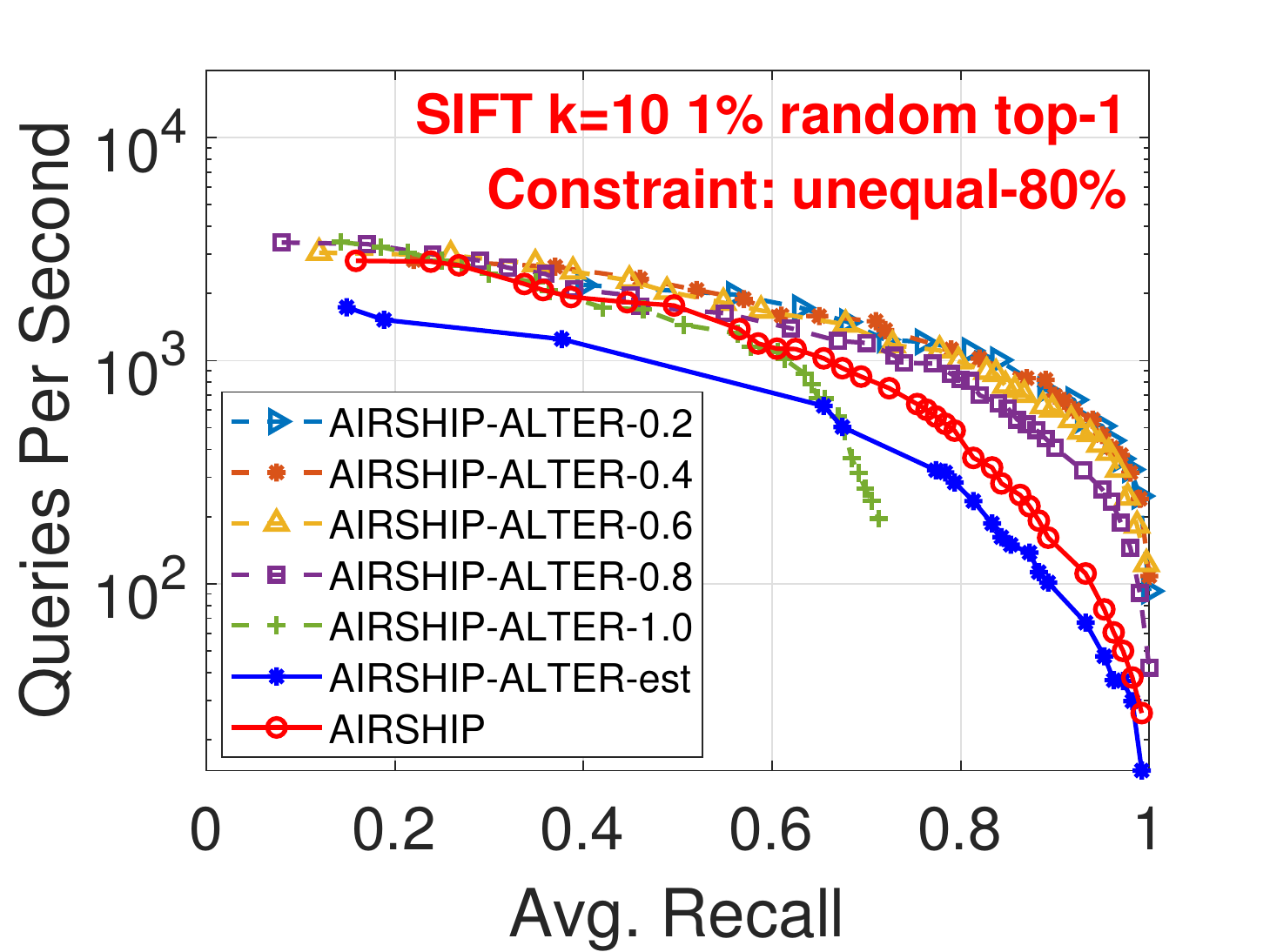}
}

\vspace{-0.14in}

\mbox{
\includegraphics[width=2.45in]{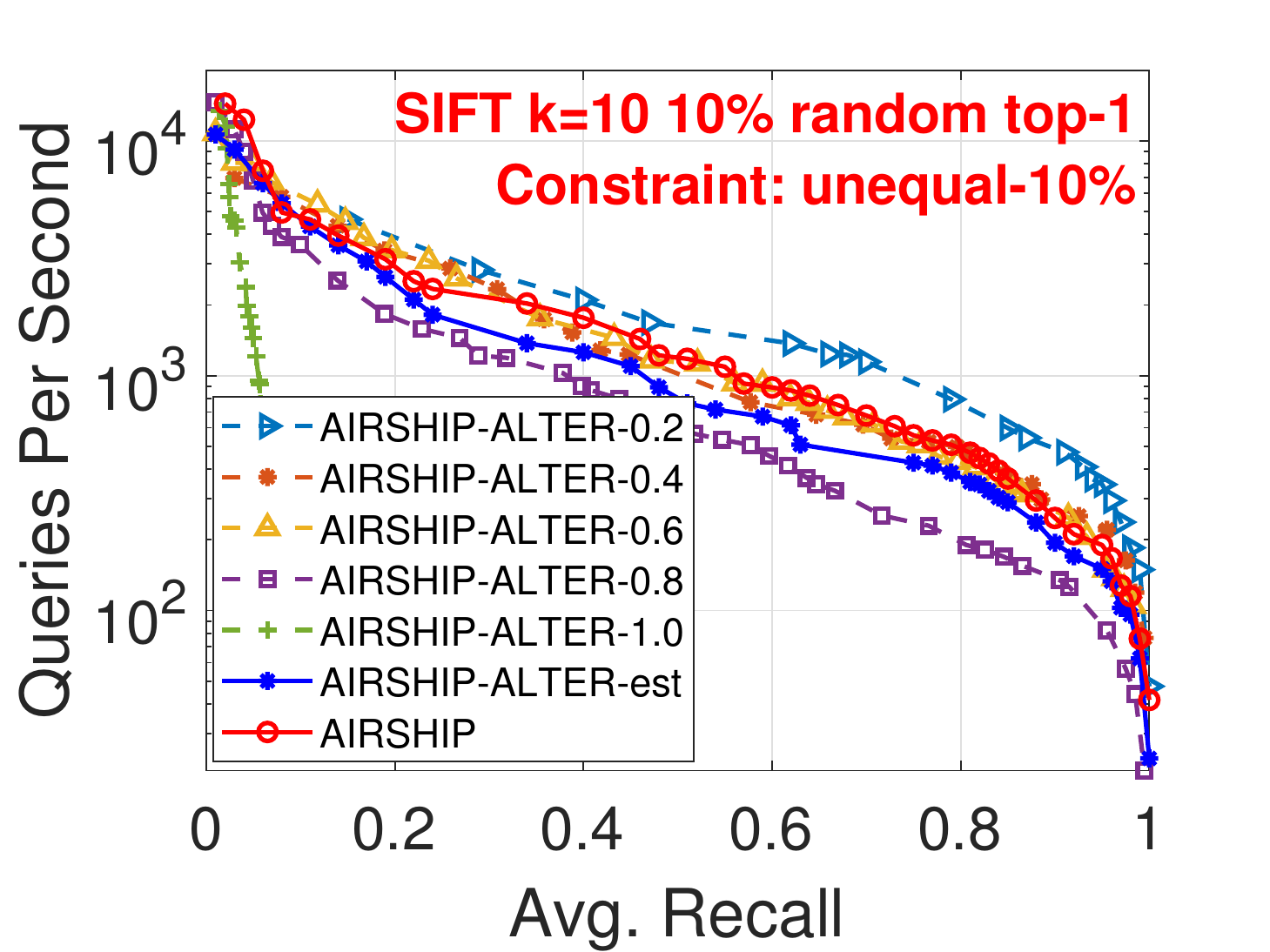}
\hspace{0.12in}
    \includegraphics[width=2.45in]{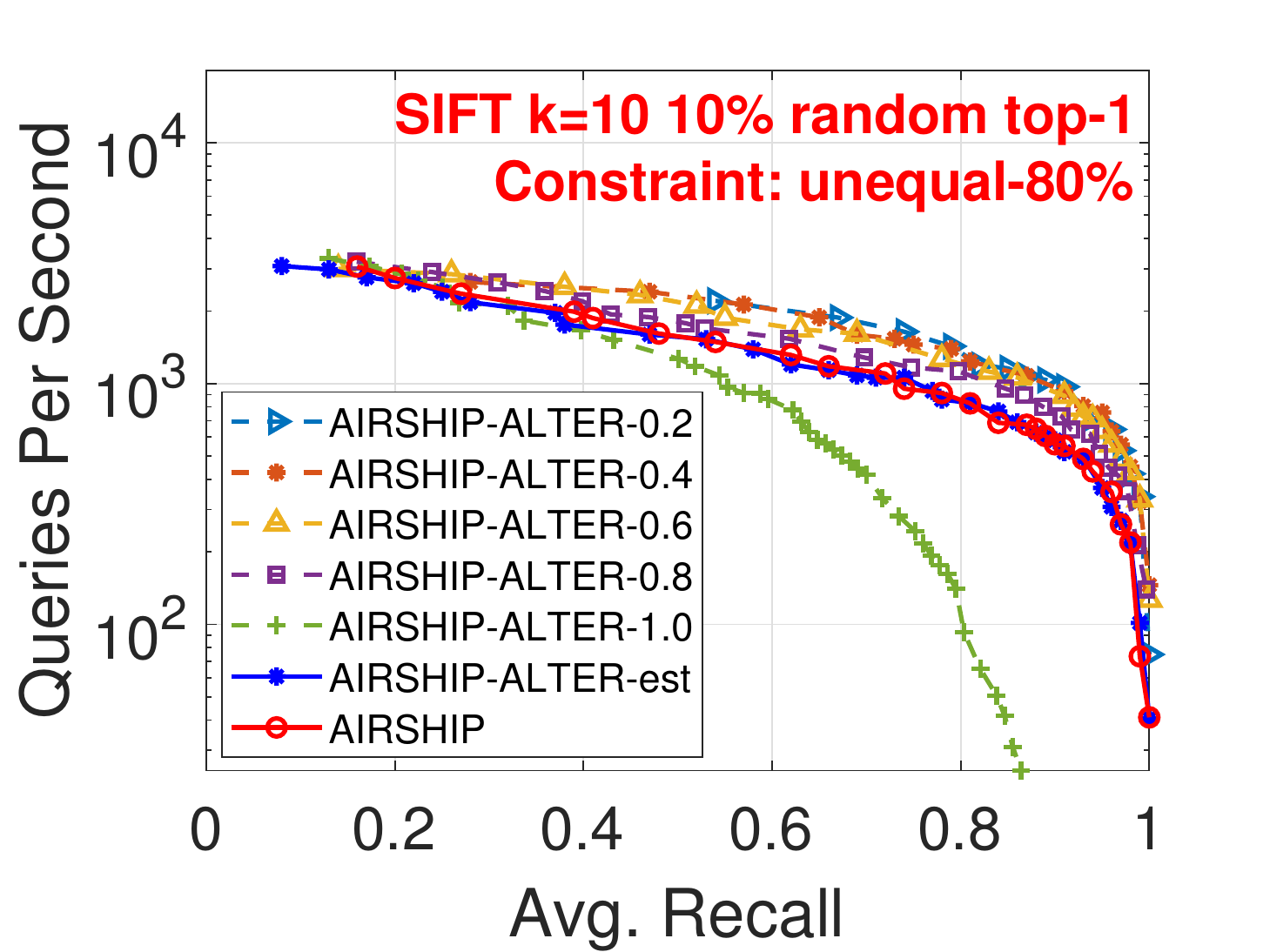}
}

\vspace{-0.14in}

\mbox{
\includegraphics[width=2.45in]{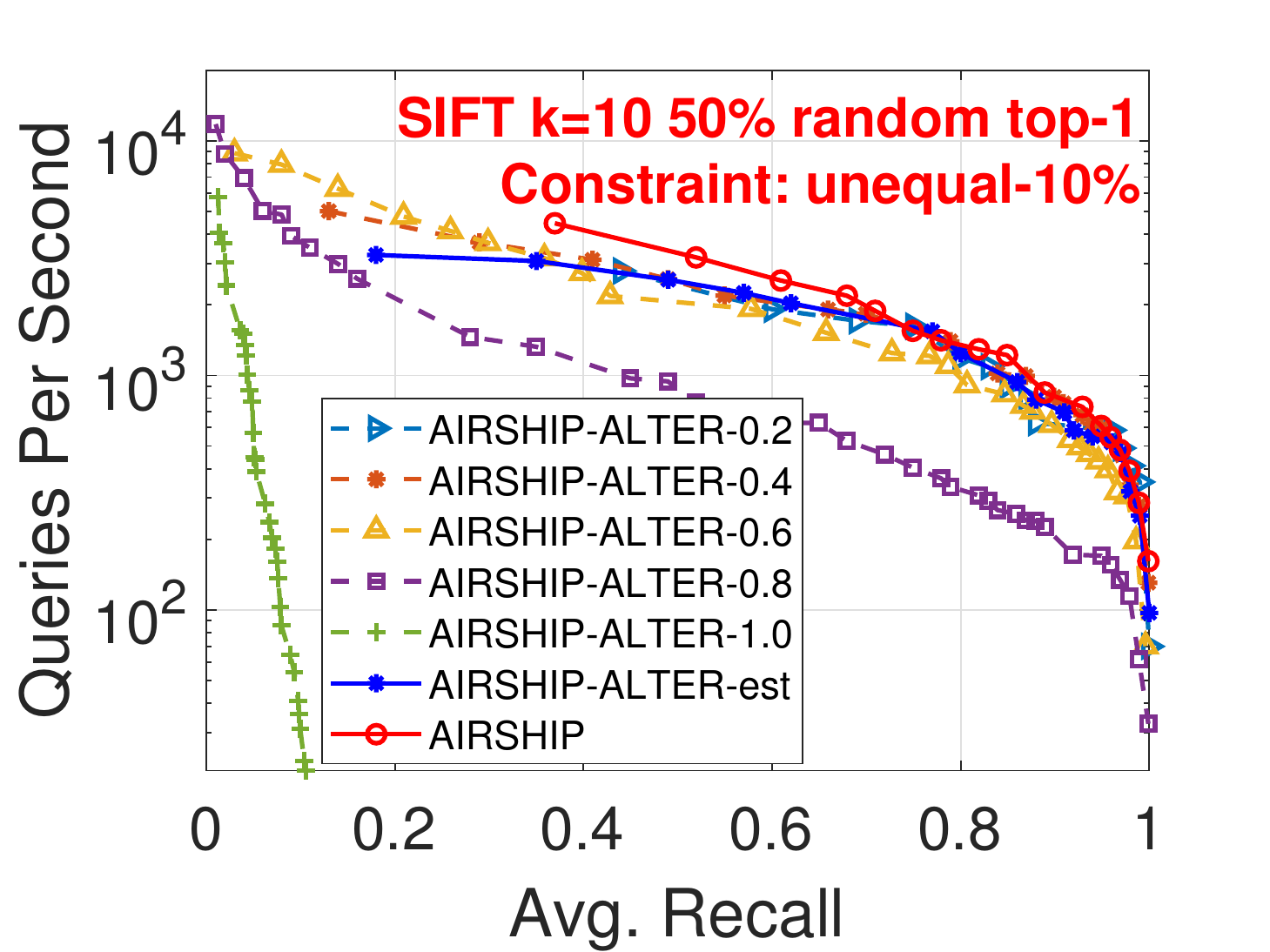}
\hspace{0.12in}
    \includegraphics[width=2.45in]{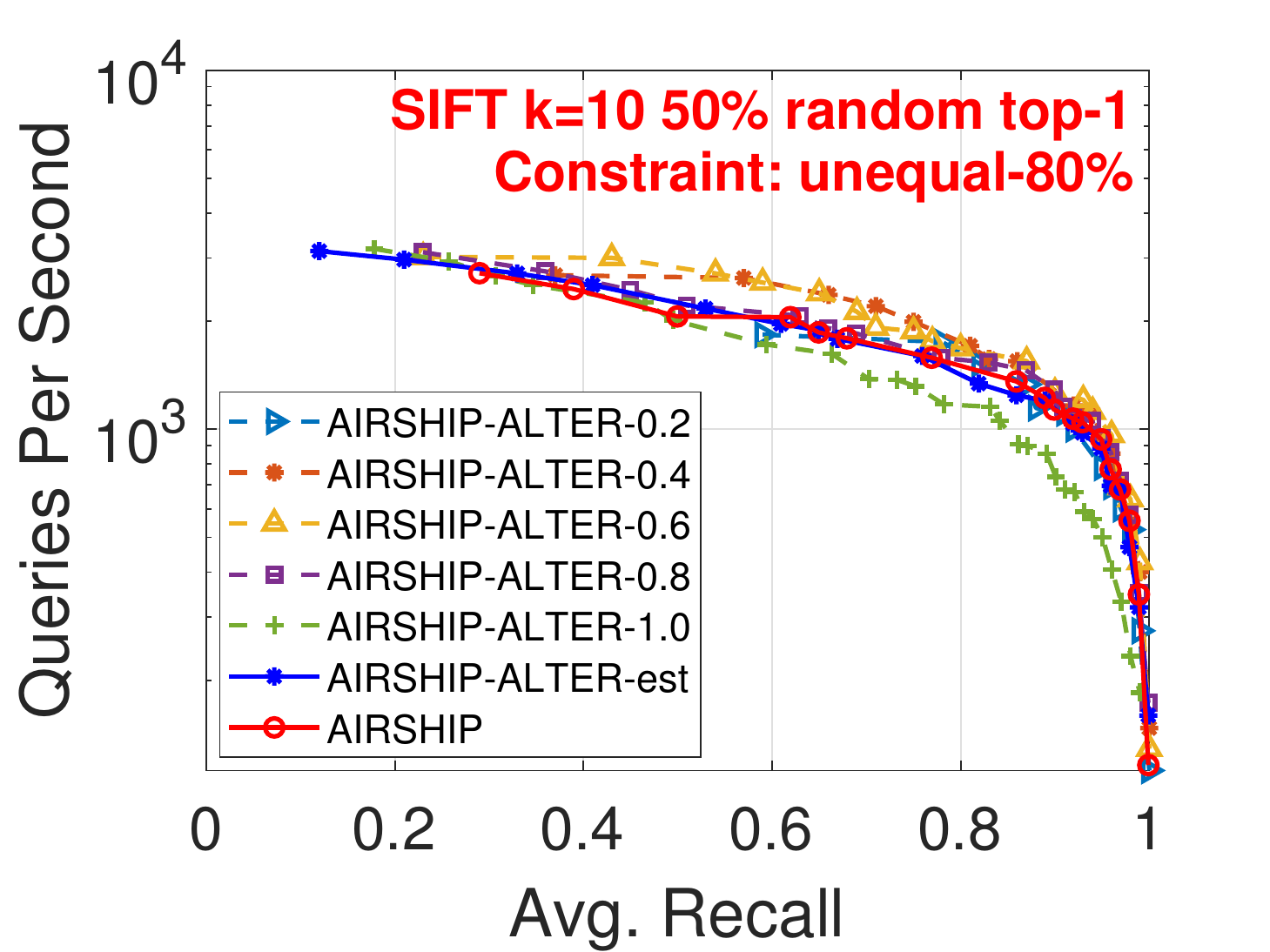}
    }

\vspace{-0.14in}

\mbox{
\includegraphics[width=2.45in]{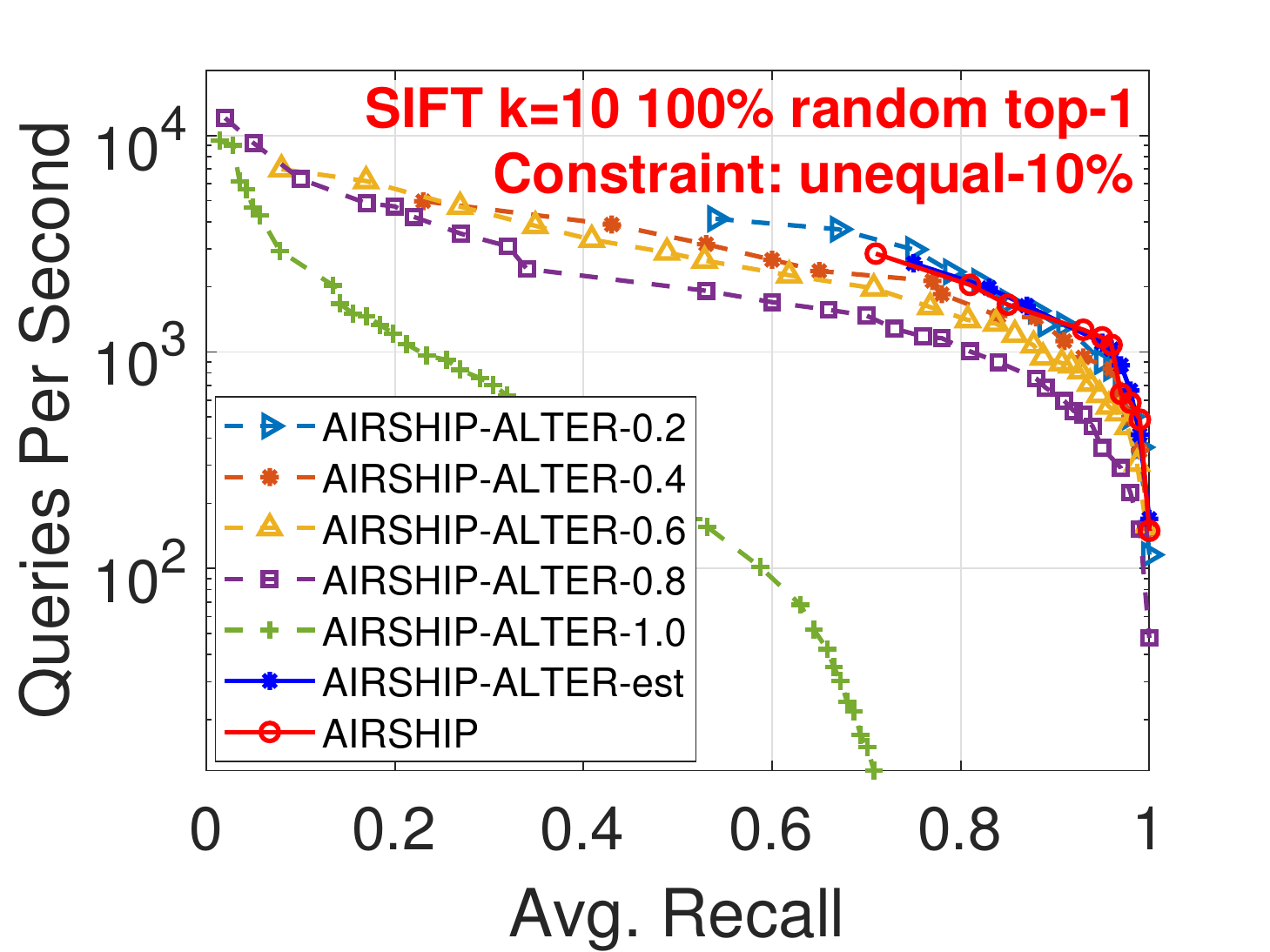}
\hspace{0.12in}
    \includegraphics[width=2.45in]{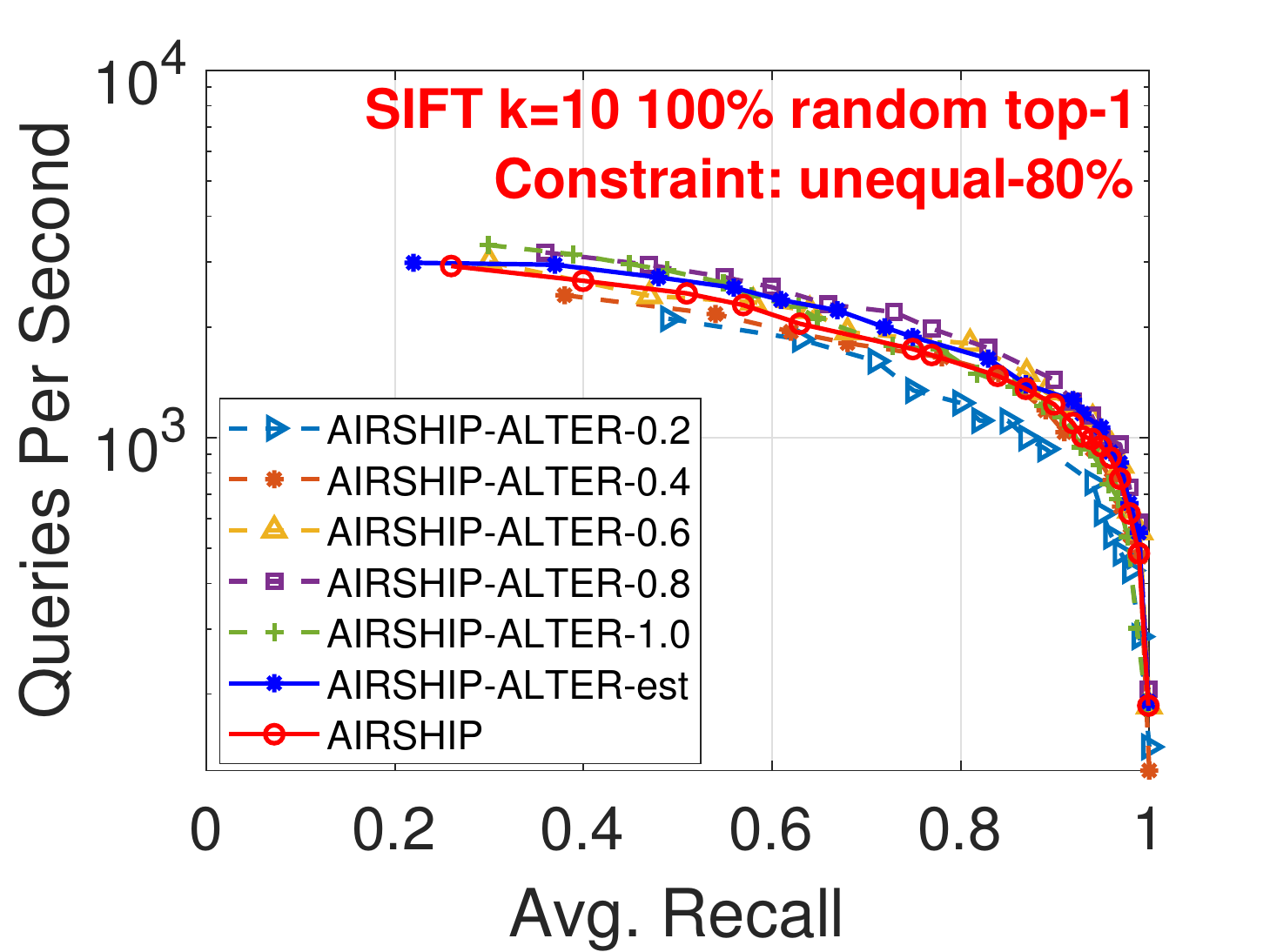}
    }
\end{center}

\vspace{-0.2in}

 \caption{Results for alpha ratio estimation on SIFT over  different amounts of random labels.
}\label{fig:exp-alter-ratio}
 \vspace{-0.2in}
\end{figure}

\newpage\clearpage

\subsection{Optimization Effectiveness}\label{ssec:exp-prefer}

\vspace{0.1in}
\noindent
\textbf{\textit{alter\_ratio} estimation.}
We evaluate the estimation of our \textit{alter\_ratio} in Figure~\ref{fig:exp-alter-ratio} by comparing the searching performance using the estimated ratio (AIRSHIP-ALTER-est) with methods using constant \textit{alter\_ratio}. We enumerate the ratio in 0.2, 0.4, 0.6, 0.8, and 1.0. The comparison is conducted on 3 different label randomness data (0\%, 1\%, 10\%, 50\%, and 100\%) and 2 query constraints (\texttt{unequal-10\%} and \texttt{unequal-80\%}).
For clustered data ($0\% random$), higher \textit{alter\_ratio} yields better performance. When the randomness of the label goes upper, i.e., $50\% and 100\%$, A low \textit{alter\_ratio} gains the best throughput (Queries Per Second). Our estimated \textit{alter\_ratio} (AIRSHIP-ALTER-est) shows a comparable performance to the best constant \textit{alter\_ratio}. Therefore, we can conclude that our estimation method effectively determines the ratio without having prior knowledge of the query constraint.

\vspace{0.1in}
\noindent
\textbf{Biased priority queue selection.}
We introduced an aggressive priority queue selection algorithm in Section~\ref{ssec:prefer}: When the top candidate from the satisfied vector queue is better than the one from the other queue, we override the $alter\_ratio$ restriction and select the satisfied vector queue.
We denote this optimization as AIRSHIP-Alter-Prefer or AIRSHIP for short since it consists of all the proposed optimizations. Figure~\ref{fig:exp-alter-ratio} shows that AIRSHIP obtains the best performance in most randomness and query constraint combinations. For $50\%$ and $100\%$ label randomness cases, it was slightly slower than AIRSHIP-ALTER-est since the biased queue selection is designed for the clustered satisfied vectors.

\begin{figure}[b!]
\vspace{-0.2in}
\begin{center}
 \mbox{
\includegraphics[width=2.45in]{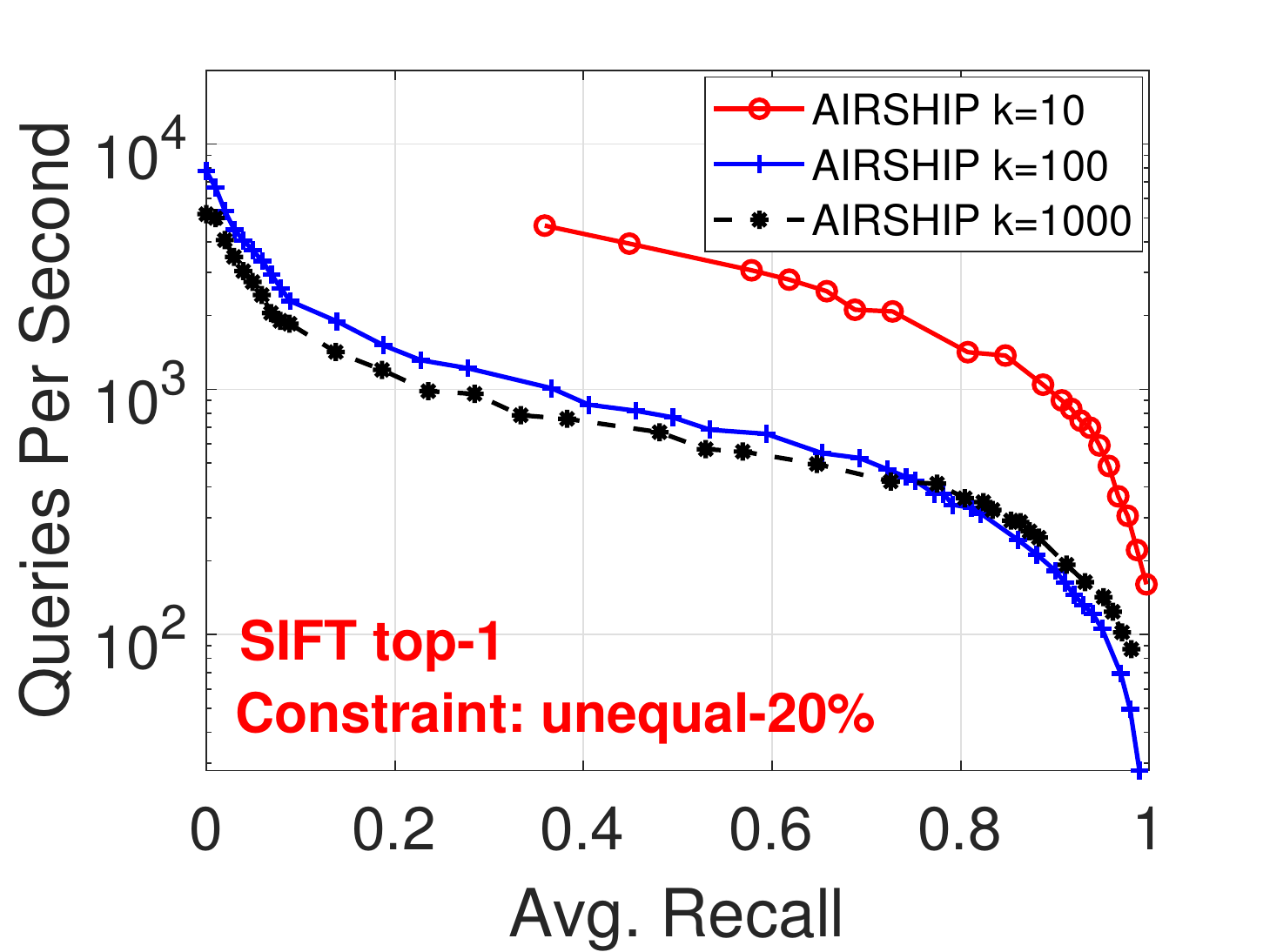}
    \hspace{0.15in}
    \includegraphics[width=2.45in]{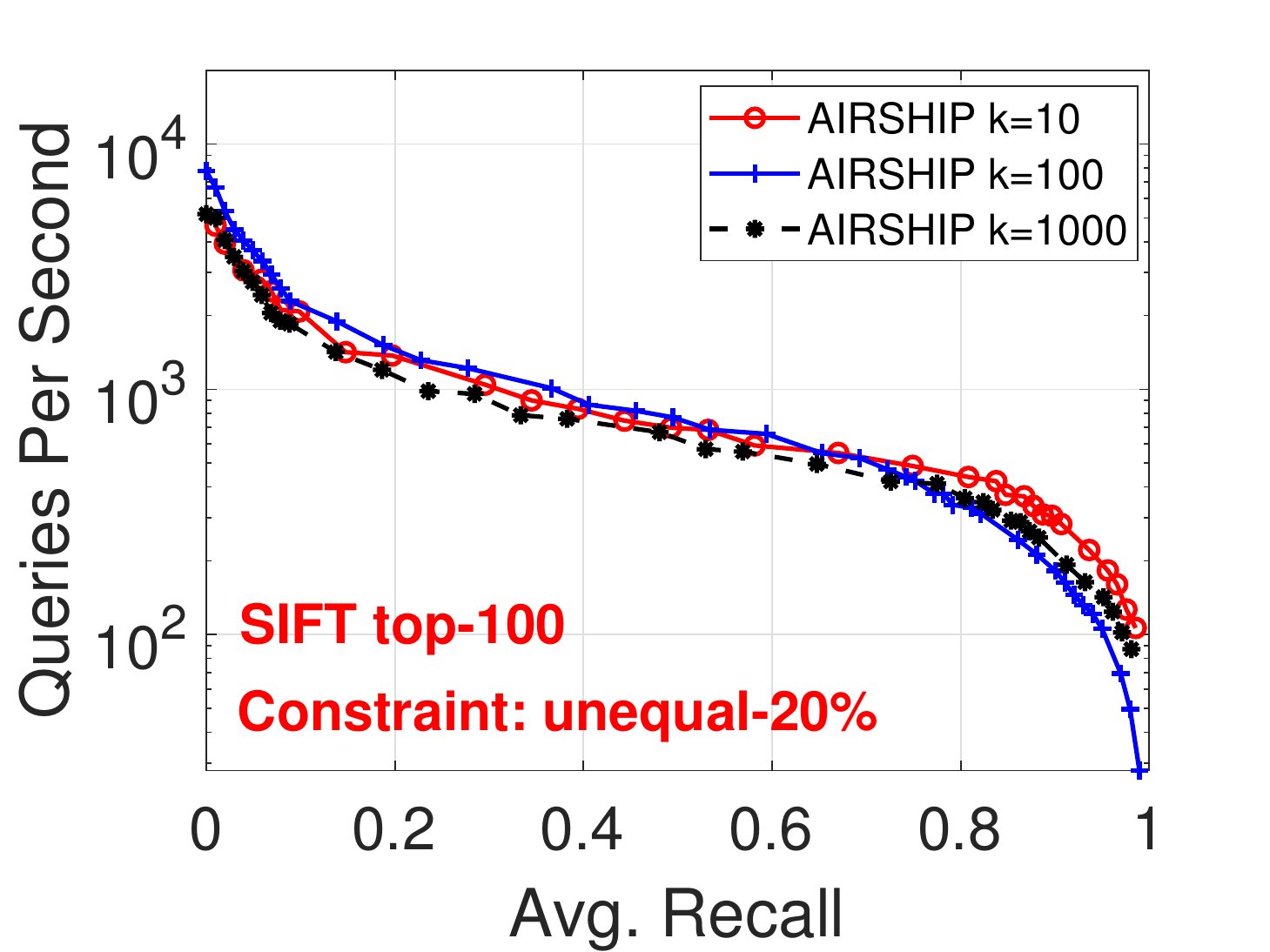}
}

\mbox{
\includegraphics[width=2.45in]{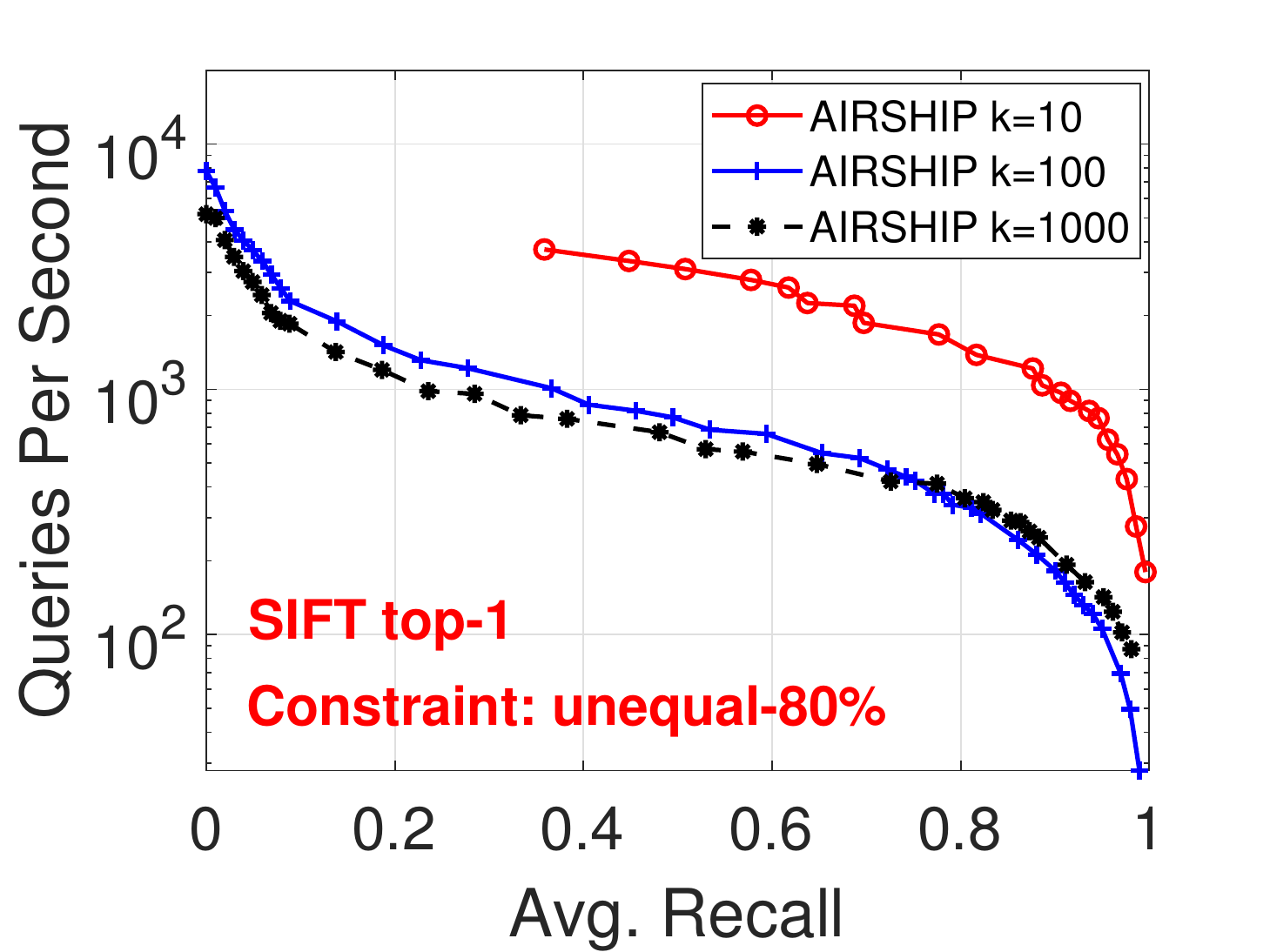}
    \hspace{0.15in}
    \includegraphics[width=2.45in]{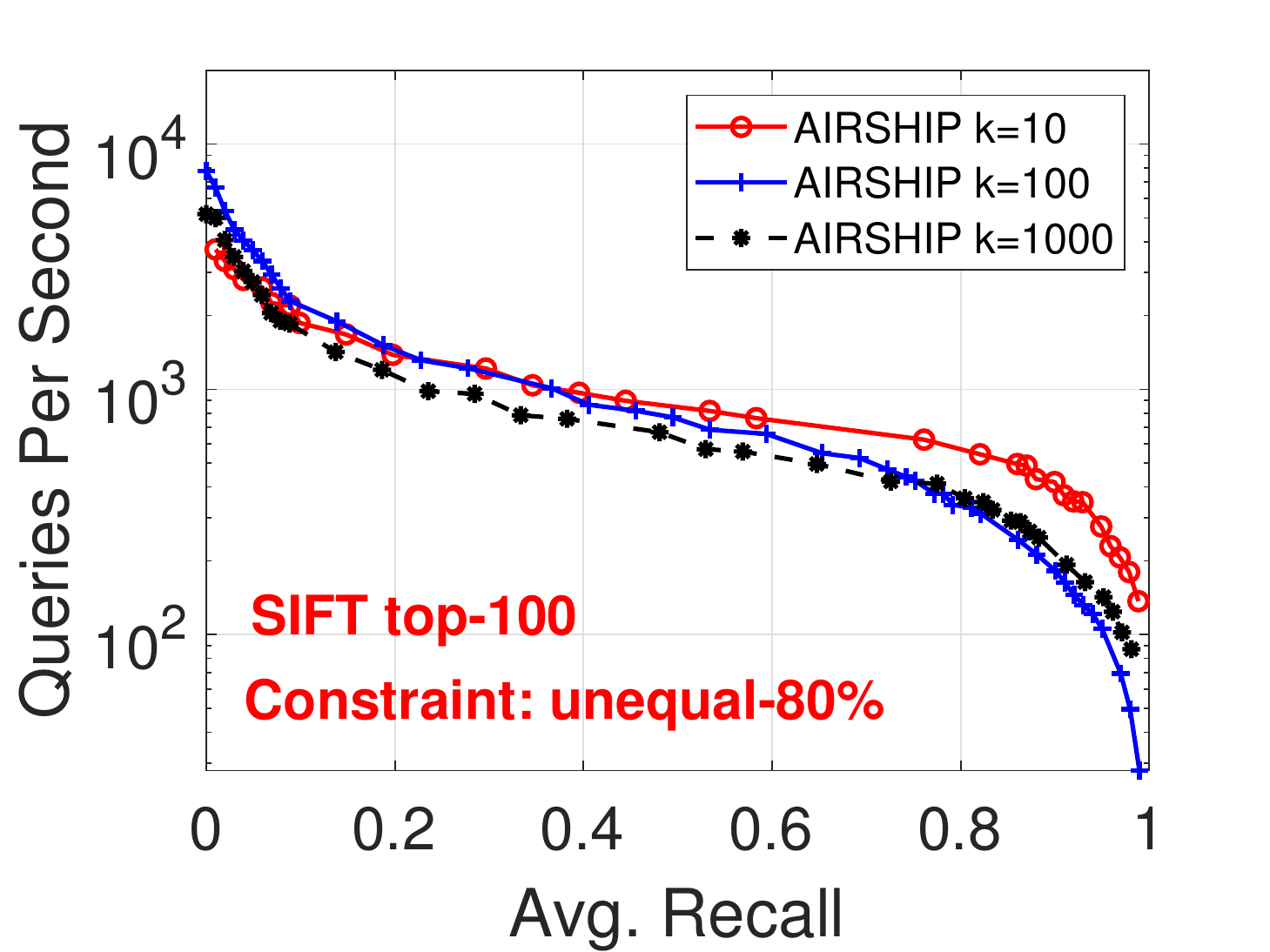}
}

\end{center}
\vspace{-0.25in}
 \caption{Results for varying the number of clusters, k.
}\label{fig:vary-k}
 \vspace{-0.15in}
\end{figure}

\subsection{Number of Satisfied Vector Clusters}
In previous experiments, our dataset has $k=10$ distinct labels that correspond to $k$ clusters of satisfied vectors. It can be viewed as the dataset contains $k$ classes. In Figure~\ref{fig:vary-k}, we report a comparison when we vary $k$ for 10, 100, and 1,000 distinct labels. The top-1 results illustrate more diverged curves than the top-100 ones because top-1 is more sensitive---we have to find the exact most similar satisfied vector. AIRSHIP has a better performance when the number of labels (classes) is small for top-1 situation, because there are fewer satisfied vector clusters to explore. The difference disappears when we consider top-100 results, since the expected numbers of vectors to be explored are the same for all three cases---the top-100 candidates are distributed across the same proportion ($20\%$ or $80\%$) of satisfied clusters.

\subsection{MNIST}

\begin{figure} [h]
\vspace{-0.1in}

 \mbox{\hspace{-0.2in}
\includegraphics[width=2.4in]{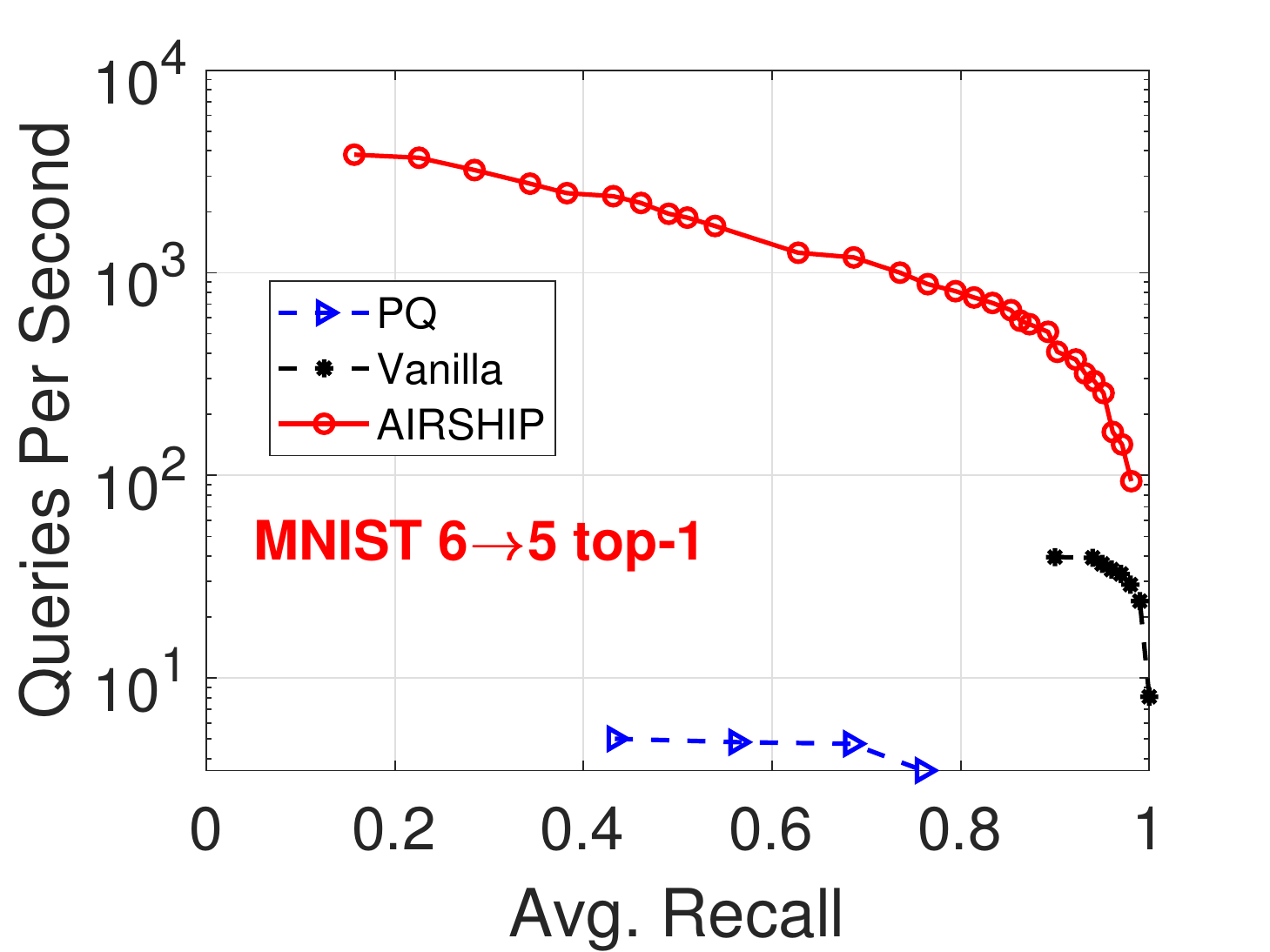}
    \hspace{-0.15in}
    \includegraphics[width=2.4in]{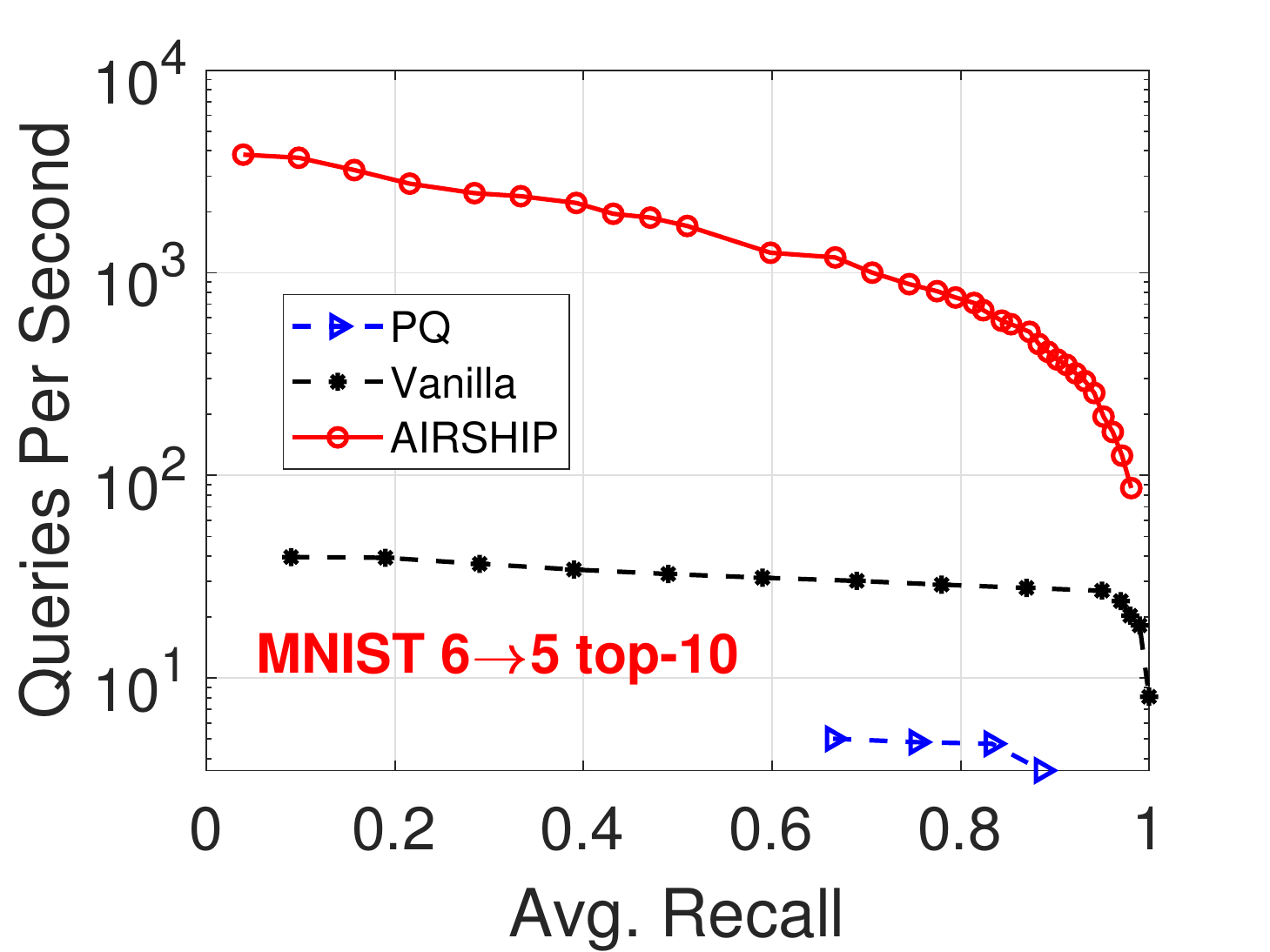}
    \hspace{-0.15in}
    \includegraphics[width=2.4in]{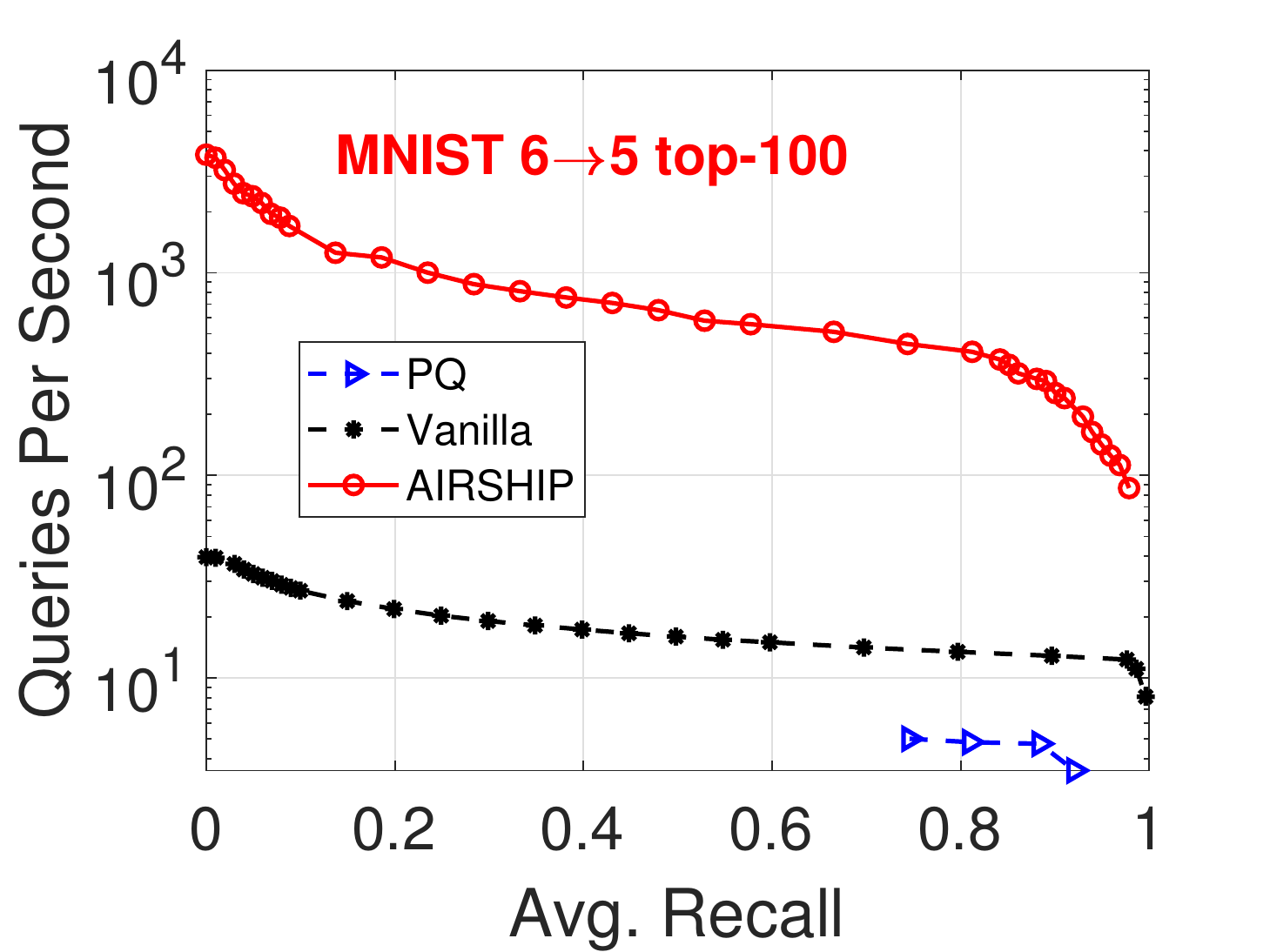}

}

 \mbox{\hspace{-0.2in}
    \includegraphics[width=2.4in]{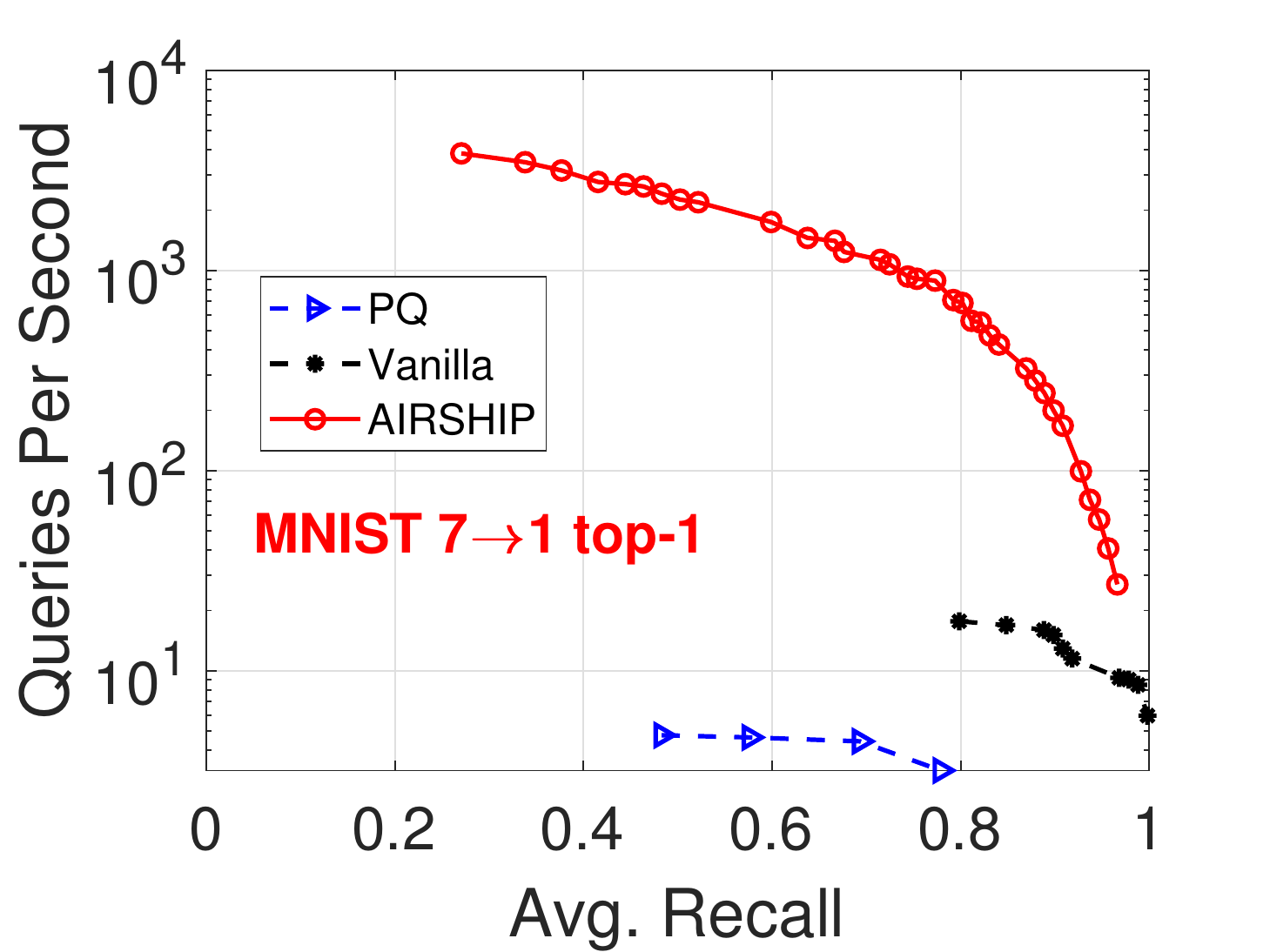}
    \hspace{-0.15in}
    \includegraphics[width=2.4in]{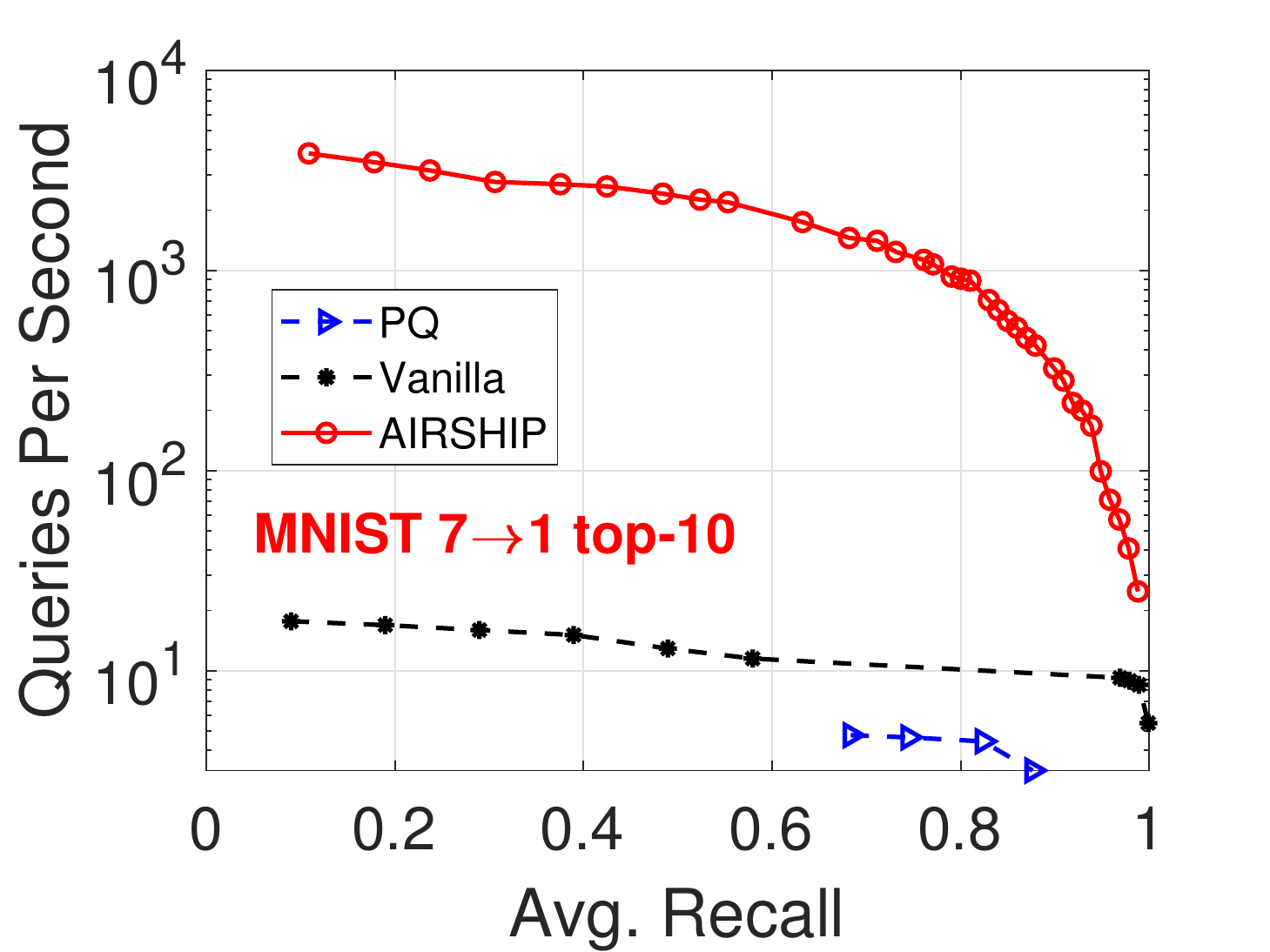}
    \hspace{-0.15in}
    \includegraphics[width=2.4in]{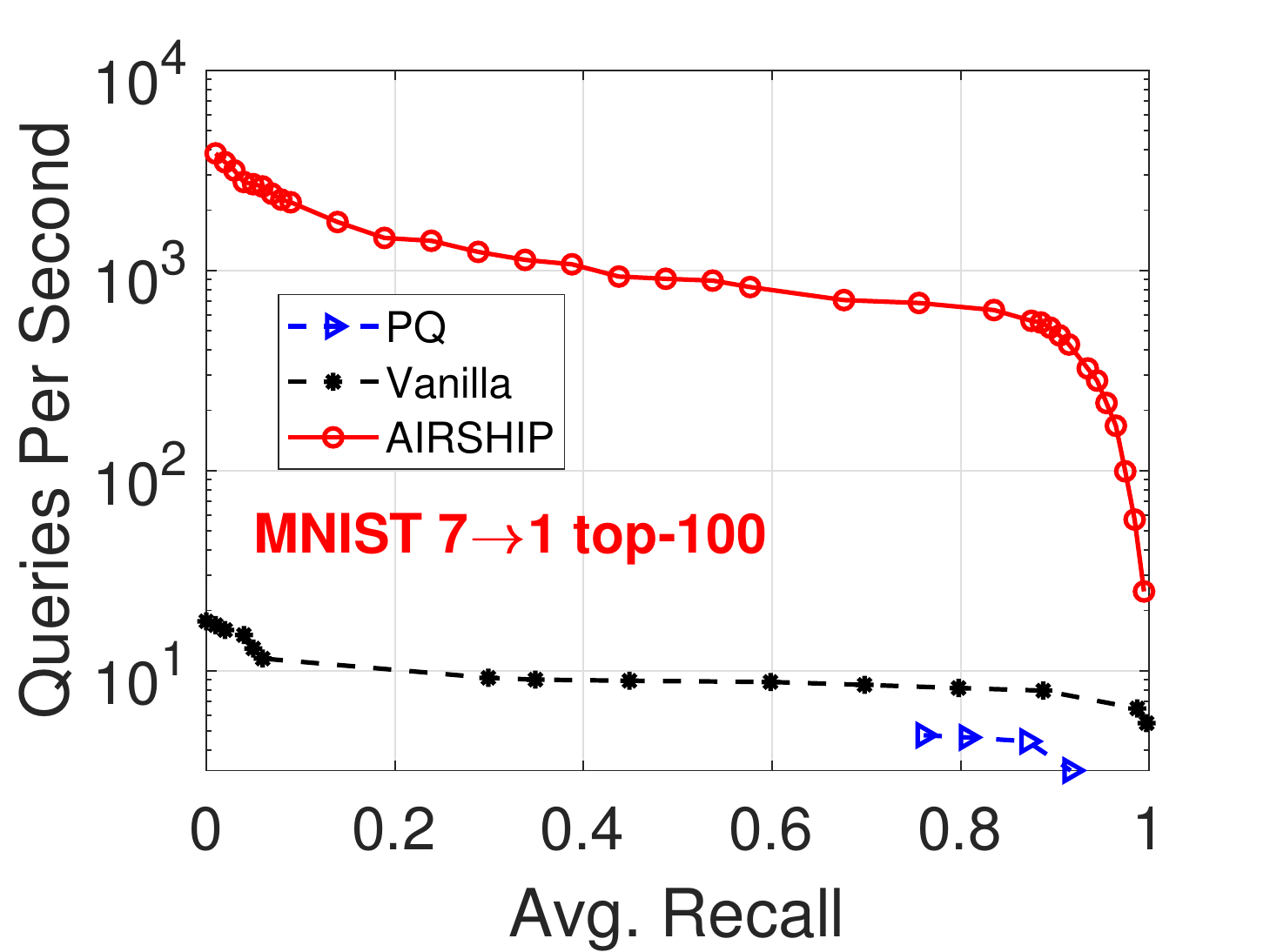}
}

\vspace{-0.1in}

 \caption{Results on MNIST. (1) searching ``5'' by ``6'' ; (2) searching ``1'' by ``7''.
}\label{fig:mnist}  
\end{figure}

\begin{figure}[b!]

\vspace{-0.2in}
\begin{center}
\includegraphics[width=0.6\textwidth]{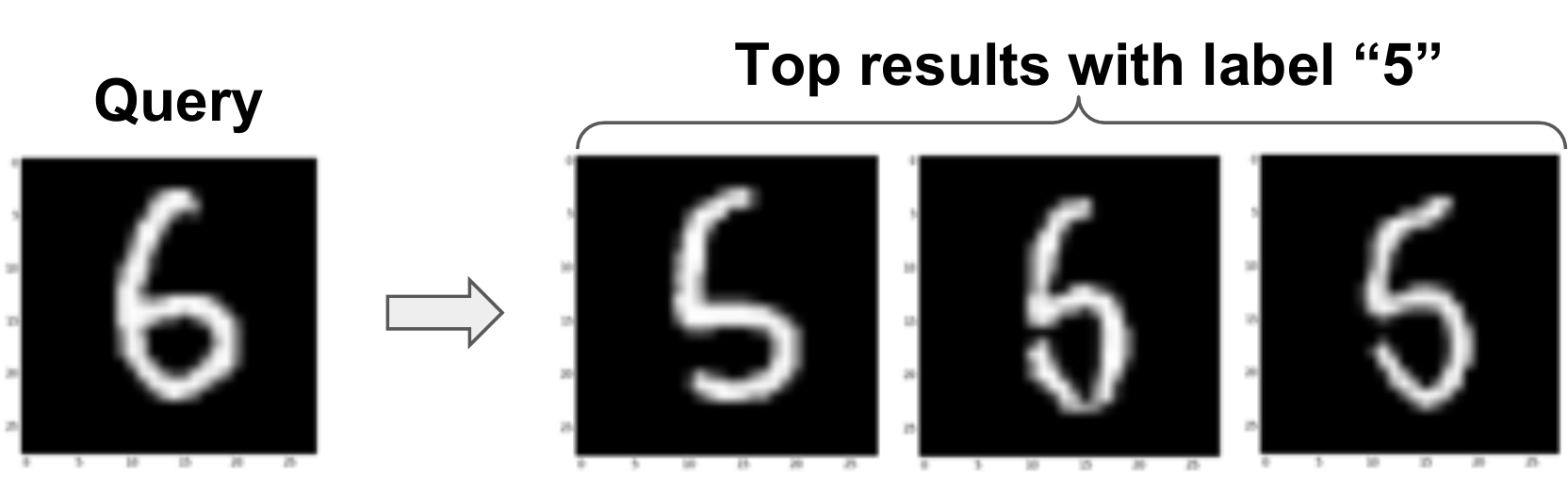}
\end{center}
\vspace{-0.15in}

\caption{An Example of searching ``5'' by ``6'' on MNIST.}
\label{fig:6to5}\vspace{-0.15in}
\end{figure}

MNIST is a hand-written digits datasets. Each vector in MNIST has its label (10 classes from 0 to 9). We use MNIST as an experiment to check the performance of AIRSHIP on real-world data distribution. Figure~\ref{fig:mnist} shows a performance comparison with two queries. (a) $6\rightarrow5$: the query point is from class $6$ and the returned similar vectors must be class $5$; (b) $7 \rightarrow 1$: given a query vector from class $7$ and find the most similar vector from class $1$. PQ suffers from computing the distances between all satisfied vectors ($10\%$). AIRSHIP achieves around 100X speedup over the vanilla graph search. The speedup is consistent across top-1, 10, and 100 results. Figure~\ref{fig:6to5} visualizes one query image and its top-3 retrieved candidates for the $6 \rightarrow 5$ constraint. This kind of constrained query can be employed not only in the recommendation system but also in the security context and negative sampling finding tasks. Although this task seems to be simple by partitioning the base vectors according to their labels. However, there can be around $2^{10}$ different queries when we target to obtain a vector within a collection of labels---it is time-consuming and space-inefficient to construct this large number of indices. Our proposed algorithm AIRSHIP does not need to build extra indices---we just perform the similarity search on the original proximity graph without worrying about the unknown query constraints.

\subsection{Discussion}
Based on the results reported above, we can answer the questions driving our experimental evaluation.
AIRSHIP substantially improves over other similarity search algorithms, i.e., HNSW, PQ. The speedup can be 10-100X depending on the data distributions.
The starting point selection optimization (AIRSHIP-Start) slightly improves over the vanilla graph search algorithm. The multi-direction priority queue search (AIRSHIP-Alter) is the major source of the speedup. The biased priority queue selection is beneficial when the satisfied labels are highly clustered.
Our estimation to \textit{alter\_ratio} on different data distributions (0\%, 1\%, 10\%, 50\%, and 100\% randomness) shows a comparable performance compared with enumerated constants. The performance of AIRSHIP is consistent across all combinations of data randomness and query constraints.

\section{Conclusion}
We investigate the \textbf{constrained similarity search} problem, and introduce \textbf{AIRSHIP}, a system that integrates a user-defined function filtering into the similarity search framework. The proposed system does not need to build extra indices nor require prior knowledge of the query constraints.
We propose three optimizations for the constrained searching problem: starting point selection, multi-direction search, and biased priority queue selection. All three optimizations improve the similarity search performance. The proposed algorithm has a hyper-parameter \textit{alter\_ratio} that is data/query dependent. We present an estimation method to adaptively choose this hyper-parameter.
Experimentally evaluations of the proposed algorithm on both synthetic and real-world data confirm the effectiveness of our optimizations: AIRSHIP is 10-100X faster than baseline solutions.\\

\noindent The constrained similarity search problem arises from many practical applications in search and advertising when the authors worked with product teams.  Obvious solutions in some cases did work well but there are many practical scenarios which required more effective solutions. Evaluating the constraints can be very expensive unless they are properly indexed. When engineers first apply ANN on the original data and apply filtering on the remaining data vectors, they often have to substantially increase the number vectors returned for ANN (e.g., returning 10000 vectors while the target is to find top-1000 vectors which also satisfy the constraints), leading to substantial performance loss. Our proposed solution has been deployed for production by the ads team, The empirical evaluations presented in this paper, however, are solely based on public datasets. \\

\noindent We choose to develop solutions for the constrained similarity search problem, based on graph-based ANN. This is because graph-based ANN is in the production system owing to its excellent performance. Among graph-based ANN, hashing-based ANN, and quantization-based ANN algorithms, typically hashing-based ANN is the simplest in terms of indexing building and the performance is acceptable if a proper hashing function is chosen~\citep{shrivastava2014defense}. Nevertheless, if the goal is to optimize for high efficiency (even at the cost of building extra indexing for graphs), then graph-based ANN should be the choice. We expect that researchers and engineers, with non-trivial effects, can also develop solutions for constrained similarity search based on other ANN techniques.

\newpage\clearpage

\bibliographystyle{plainnat}
\bibliography{refs_scholar}

\end{document}